\documentclass[12pt]{iopart}

\usepackage{graphicx}
\usepackage{subfigure}
\usepackage{url,graphics,epsfig}
\usepackage{amssymb}
\usepackage{color}
\begin{document}

\title[Towards Quantum Simulation of Chemical Dynamics]{Towards Quantum Simulation of Chemical Dynamics with Prethreshold Superconducting
Qubits} 

\author{P. C. Stancil$^{1,2}$, H. You$^{1,2}$, A. Cook$^{1,2}$,
A. T. Sornborger$^{3}$, and M. R. Geller$^{1}$}

\address{$^{1}$Department of Physics and Astronomy,\\
                        University of Georgia, Athens, GA 30602-2451, USA}
\address{$^{2}$Center for Simulational Physics,\\
                        University of Georgia, Athens, GA 30602-2451, USA}
\address{$^{3}$Department of Mathematics, University of California, Davis, CA 95616, USA}

\ead{stancil@physast.uga.edu}
\vspace{10pt}
\begin{indented}
\item[]\today
\end{indented}

\begin{abstract}
The single excitation subspace (SES) method for universal quantum simulation is investigated for
a number of diatomic molecular collision complexes. Assuming a system of $n$ tunably-coupled, and fully-connected superconducting qubits, computations
are performed in the $n$-dimensional SES which maps directly to an $n$-channel collision problem within
a diabatic molecular wave function representation. Here we outline the approach on a classical computer
to solve the time-dependent Schr\"odinger equation in an $n$-dimensional molecular basis - the so-called
semiclassical molecular-orbital close-coupling (SCMOCC) method  - and extend the treatment beyond the
straight-line, constant-velocity approximation which is restricted to large kinetic energies ($\gtrsim 0.1$ keV/u). We explore
various multichannel potential averaging schemes and an Ehrenfest symmetrization approach to allow for
the application of the SCMOCC method to much lower collision energies (approaching 1 eV/u). In addition, a computational
efficiency study for various propagators is performed to speed-up the calculations on classical computers.
These computations are repeated for the simulation of the SES approach assuming typical parameters for
realistic pretheshold superconducting  quantum computing hardware. 
The feasibility of applying future SES processors to the quantum dynamics of large molecular collision
systems is briefly discussed. 
\end{abstract}

%
%
%
%
%

\section{Introduction}

While the field of chemical dynamics, including atomic and molecular collisional processes, has seen tremendous
advances in theory, experiment, and computation over the past eight decades \cite{alt03,bow11,nym13}, computations  which
attempt to exactly solve the time-dependent (TD) or time-independent (TI) Schr\"odinger equation have been limited
to consideration of only five \cite{wan06} and four atoms \cite{yan15}, respectively. In the former case, solutions have been restricted
to reactive processes, while the latter approach has focused on inelastic collisions, both incorporating full-dimensional dynamics on full-dimensional potential energy surfaces (PESs). Part of the reason for the dimensional limitation in such calculations is the need for significant
computational resources, but more importantly the development of software and algorithms to treat large
multidimensional systems has stagnated.

\begin{figure}[htb]
\begin{center}
\includegraphics[scale=0.4]{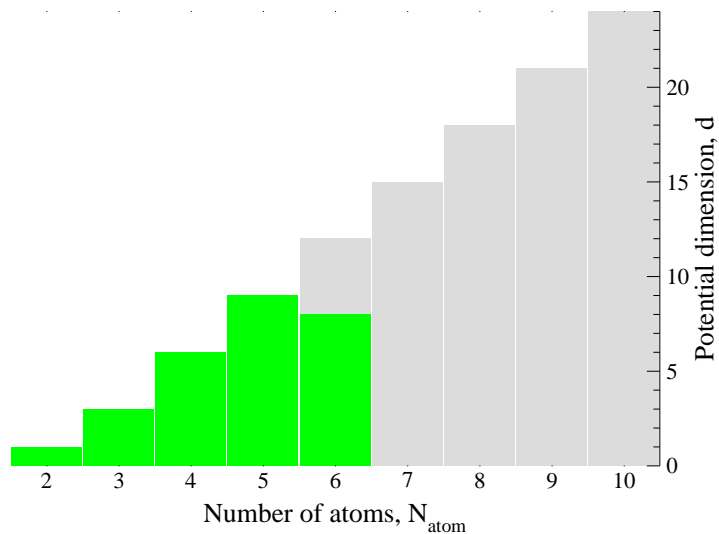}
\caption{Status of time-dependent reactive calculations on classical computers in terms of the number of
atoms $N_{\rm atom}$ and PES dimension $d$. Regime of possible quantum scattering 
calculations (green). Regime of available PESs which can currently only be treated by quasi-classical
dynamical methods (grey). }
\label{dofpot}
\end{center}
\end{figure}

On the other hand, molecular electronic structure  computations, or quantum chemistry, has seen rapid
advances in both software and algorithm development with implementation on high performance distributed
and shared-memory CPUs \cite{bow11} as well as on new accelerator technologies such as graphical processing units (GPUs)
\cite{oli10}.
Using the coupled-cluster singles and doubles (CCSD) approach \cite{molpro}, it is possible today to compute and analytically
fit electronic potential and coupling surfaces for systems as large as 10 atoms \cite{bra09}. As the number of
internal degrees of freedom $d$ is $3N_{\rm atom}-6$, where $N_{\rm atom}$ is the number of atoms, this corresponds
to a 24-dimensional surface. As an illustration, Figure~\ref{dofpot} plots the PES dimension $d$ versus
the number of atoms  up to $N_{\rm atom}=10$, which is approximately the limit for the size of the
largest molecular systems that can be both computed on a classical compute and fitted for dynamical
studies. However, the region show in green displays the largest systems that can be computed,
again on a classical computer, using full dimensional dynamics for TD reactive collisions.  The
region with $d>9$ can only be treated dynamically with quasi-classical methods, i.e. by solving
Newton's equations of motion for the heavy-particle trajectories. The situation for TI and quantum inelastic
calculations is somewhat worse. 

Clearly computations of electronic structure have far out-paced the abilities of
quantum dynamical calculations when the goal is to treat the problem nearly exact
numerically and in full-dimension. Therefore, it appears that there is an opportunity to
apply other, more novel approaches to advance quantum chemical dynamical studies
and one might naturally turn to quantum computing/simulation. There has been
considerable effort to explore the prospects of applying quantum simulation
to the electronic structure problem \cite{asp05,lan10,lu11}, but investigations of chemical
dynamics have been sparse to date \cite{lid99,smi07,kas08}. The promising method
of Kassal et al \cite{kas08} applies a quantum gate-based logic approach, or digital
quantum simulation (DQS). However, the DQS method requires  100s of gate operations
and 100s of high-fidelity, fault-tolerant qubits. As the quantum simulation hardware has
not advanced sufficiently to satisfy these resource requirements, we have proposed
an alternative approach, the single excitation subspace (SES) method, which avoids
the need for fault-tolerant devices using instead available prethreshold superconducting
technology \cite{pri10,Geller2015}. While the SES method may not be scalable, it can solve
a time-dependent, real, symmetric quantum Hamiltonian of dimension $n\times n$ using $n$ qubits 
with a quantum computation time that is independent of $n$ for a single run. However, an SES
computer must be fully connected requiring $n(n-1)/2$ tunable couplers with $n$ also
corresponding to the number of diabatic molecular channels in the collision problem to
be simulated. The feasibility of the
approach was outlined in Geller et al \cite{Geller2015} for the simple, but well-studied
$n=3$ Na+He electronic excitation problem. Here we extend that study to i) ion-atom charge
exchange systems with $n$ as large as ten, ii) improve the trajectory calculation from
the standard straight line, constant velocity approximation to explicitly solving for the relative
velocity for a range of multichannel potential averaging schemes, and iii) apply the Ehrenfest
symmetrization approach to correct for the loss of detail-balance due to potential averaging,
with the latter two topics allowing for the classical calculations and SES simulations to
be extended to low collision energies. iv) To allow for a future classical-quantum resource comparison,
a TD propagator study is carried out to find the most efficient classical computational approach
and v) we end by speculating on the prospects of large-scale SES device applications to
large, chemically interesting reactions not feasible on today's high performance computing
platforms.

\section{Molecular Collisions on a Classical Computer: Establishing Benchmarks}

While there are a variety of approaches to attack atomic and molecular collision
problems on a classical computer, the one that is most relevant to the SES method is the semiclassical
molecular-orbital close-coupling (SCMOCC) approach. In the SCMOCC method, the
TD Schr\"odinger equation is given by
\begin{equation}
i \frac{\partial \psi (\vec{r},t|R)}{\partial t} = h(\vec{r},t|R) \psi (\vec{r},t|R),
\end{equation}
where $h(\vec{r},t|R)$ is the system Hamiltonian, $R$ the internuclear distance,  $\vec{r}$ the collection
of electronic (internal) coordinates, and $t$ the collision
time \cite{child84}. The Hamiltonian is given by
\begin{equation}
h(\vec{r},t|R) = H_{\rm int}(\vec{r}) + V(R(t),\vec{r})
\end{equation}
and the system wave function $\psi(\vec{r},t|R)$ is expanded in a molecular basis by
\begin{equation}
\psi(\vec{r},t|R) = \sum_i^n a_i(t|R)\phi_i(\vec{r})
\end{equation}
with $n$ the size of the basis or total number of channels. The asymptotic states are defined by
the TI Schr\"odinger equation
\begin{equation}
H_{\rm int}(\vec{r})\phi_i(\vec{r}) = E_i \phi_i(\vec{r}) 
\end{equation}
resulting in $n$ coupled equations for the expansion coefficients
\begin{equation}
i \frac{da_i (t|R)}{d t} = \sum_j^n V_{ij}(t|R) a_j(t|R)
\end{equation}
with $V_{ij}$ the potential matrix in the basis of states $\phi_i$
\begin{equation}
V_{ij}(t|R) = \bigl <\phi_i | V(R(t),\vec{r})|\phi_j \bigr>.
\end{equation}
In the Born-Oppenheimer approximation, $V_{ij}$ are the usual electronic potentials which
are diagonal in the adiabatic representation, and non-diagonal, but real and symmetric
in the diabatic representation applied here.
 
In a semiclassical approach, quantum probabilities are propagated with time.
For a given trajectory with initial asymptotic speed $v_0$ and impact parameter $b$
starting at a collision time $t\rightarrow -\infty$, the final
probability for a transition from initial channel $i$ to final channel $f$ 
as $t \rightarrow \infty$ is
\begin{equation}
P_{if}(v_0,b,t\rightarrow \infty) = P^\infty_{if}(v_0,b) = |a_f(t\rightarrow \infty)|^2 ,
\label{prob}
\end{equation}
with the initial condition
\begin{equation}
a_i(t\rightarrow -\infty) = \delta_{ij}
\end{equation}
for a collision system with $n$ channels. At any time, unitarity must be satisfied, so that
\begin{equation}
\sum_{j=1}^n P_{ij}(v_0,b,t) = 1.
\label{unity}
\end{equation}
After performing the propagation for a large number of impact parameters over the range
$b>0$ to $b_{\rm max}$, the integral cross section at a given initial speed $v_0$ is
given by
\begin{equation}
\sigma_{if}(v_0) = 2\pi \int^{b_{\rm max}}_0 P^\infty_{if}(v_0,b)bdb,
\label{sigsc}
\end{equation}
where for
\begin{equation}
b > b_{\rm max},  ~~~~P^\infty_{if}(v_0,b) \rightarrow 0~~~ {\rm and}~~~ P^\infty_{ii}(v_0,b) \rightarrow 1.
\end{equation}

As test cases, we expand upon our earlier atom-atom $n=3$ Na+He electronic excitation work \cite{Geller2015} 
and consider larger ($n=3-10$)  ion-atom charge exchange collisions. See Refs. \cite{Geller2015,lin08}
for details on the Na-He potential matrix and straight line trajectory calculations.

\subsection{Si$^{3+}$ + He}

\begin{figure}[htb]
\begin{center}
\includegraphics[scale=0.40]{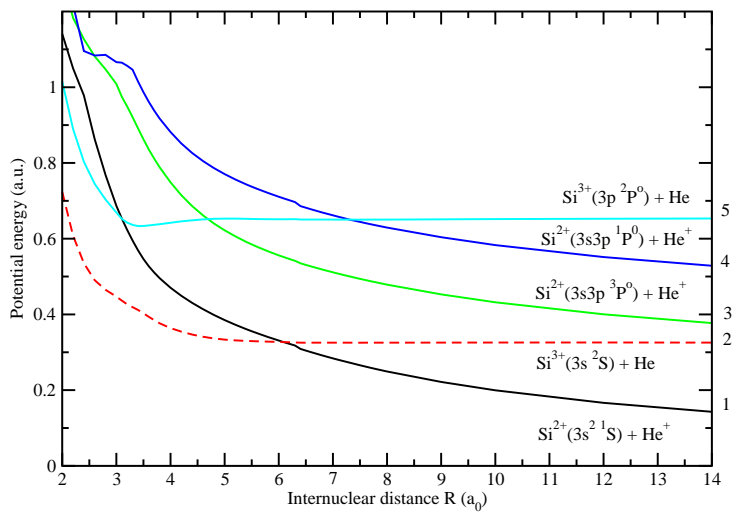}
\caption{Diabatic potentials for the SiHe$^{3+}$ system from Ref.~\cite{sta99}.
Channel numbers are indicated on the right side. The initial channel in the current simulations
is taken to be 2 or 1.}
\label{pots}
\end{center}
\end{figure}

The charge exchange process
\begin{eqnarray}
{\rm Si}^{3+}(3s~^2S) + {\rm He}(1s^2~^1S) & \rightarrow &  {\rm Si}^{2+}
(3s^2~^1S) + {\rm He}^+(1s~^2S),\\
& \rightarrow &  {\rm Si}^{2+}
(3s3p~^3P^o) + {\rm He}^+(1s~^2S), \\
& \rightarrow &  {\rm Si}^{2+}
(3s3p~^1P^o) + {\rm He}^+(1s~^2S).
\label{sihereaction}
\end{eqnarray}
was studied by Stancil et al \cite{sta99} using a TI quantum molecular-orbital close-coupling
(QMOCC) approach \cite{kim90,zyg92}. It is an $n=5$ channel case which also includes excitation to
Si$^{3+}$($3p~^2P^0$). The diabatic PESs, $V_{ii}$, are displayed in Fig.~\ref{pots},
while the off-diagonal coupling elements $V_{ij}$ can be found in Ref. \cite{sta99}.

\begin{figure}[h]
\begin{center}
\includegraphics[scale=0.4]{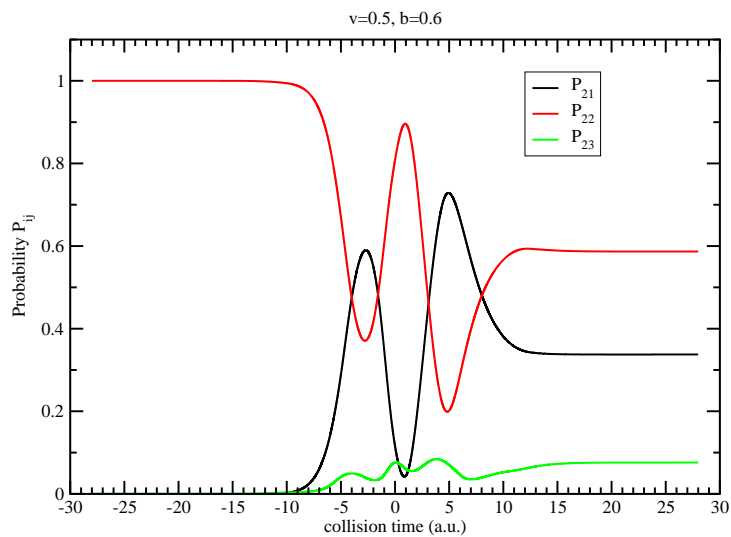}
\caption{The Si$^{3+}$ + He probabilities as a function of collision time for the elastic ($2\rightarrow 2$) and
charge exchange transitions ($2\rightarrow 1$ and $2\rightarrow 3$).
$n=3$ channel case with $b=0.6$ a$_0$ and $v_0=0.5$ a.u.}
\label{probsihe3}
\end{center}
\end{figure}

\begin{figure}[h]
\begin{center}
\includegraphics[scale=0.4]{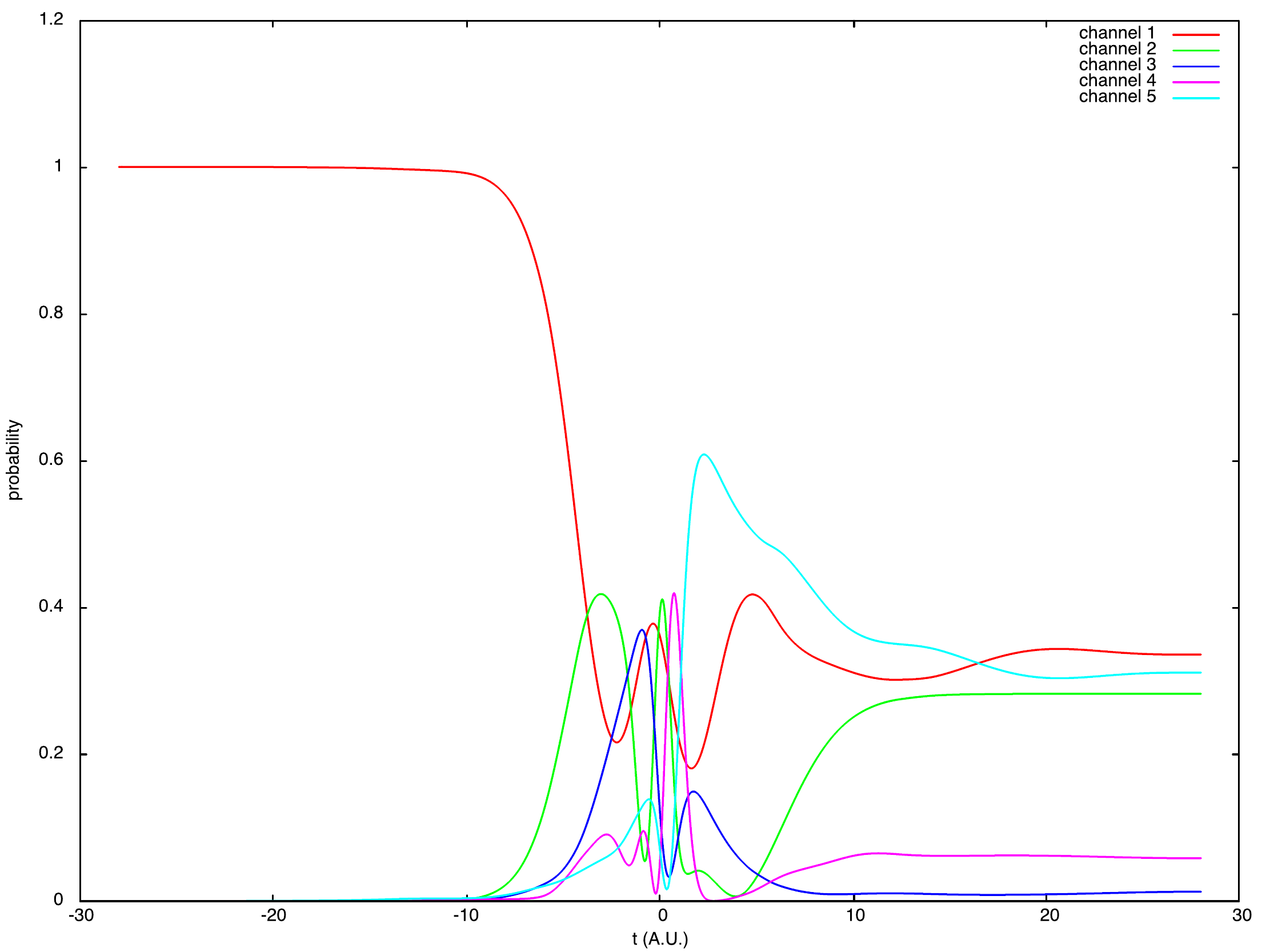}
\caption{The Si$^{3+}$ + He probabilities as a function of collision time with channel 1 being the
initial state for the elastic ($1\rightarrow 1$) and
charge exchange transitions ($1\rightarrow 2$ and 5)  and excitation ($1\rightarrow 3$ and 4).
$n=5$ channel case with $b=1.0$ a$_0$ and $v_0=0.5$.}
\label{tAtom_vs_prob}
\end{center}
\end{figure}

However, we begin by considering just the first three channels (i.e., $n=3$) with probabilities versus collision time given in Fig.~\ref{probsihe3} for
$v_0=0.5$ a.u. and $b=0.6$ a$_0$. The dominant capture channel is to the exoergic channel 1. 
A $n=5$ calculation is shown in Fig.~\ref{tAtom_vs_prob} with the ground state being the initial channel.
Other probability evolution examples for various $v_0$ and $b$ and for $n=4$ and $n=5$ simulations
are given in the Supplement.

\begin{figure}[h]
\begin{center}
\includegraphics[scale=0.4]{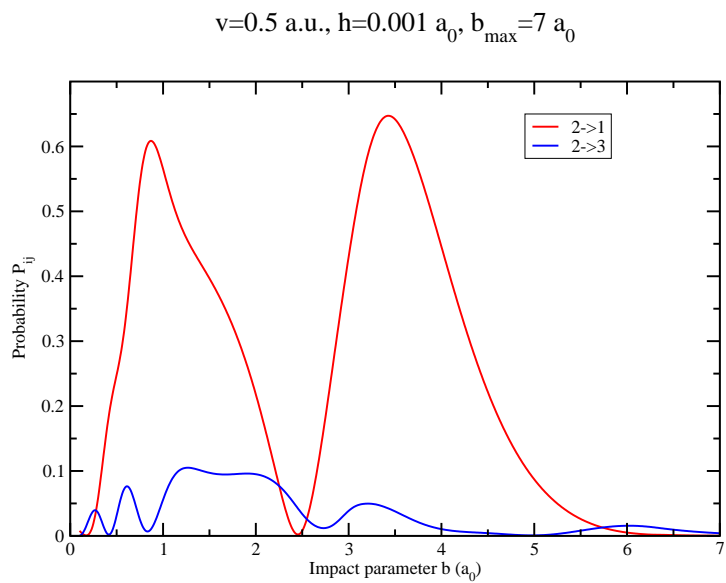}
\caption{The Si$^{3+}$ + He charge exchange probability for the $2\rightarrow 1$ ($2~^2\Sigma^+
\rightarrow 1~^2\Sigma^+$) and $2\rightarrow 3$ ($2~^2\Sigma^+
\rightarrow 3~^2\Sigma^+$) transitions versus
impact parameter. Three channel case with $v_0=0.5$ a.u.}
\label{pbsihe3}
\end{center}
\end{figure}

Figure~\ref{pbsihe3} displays the charge exchange probabilities versus $b$ for $v_0=0.5$ a.u. and $n=3$ For the
dominant $2\rightarrow 1$ transition, two main probability peaks are evident with the probability
falling off to zero by $b=7$ a$_0$. Additional examples are given in the Supplement.

\begin{figure}[h]
\begin{center}
\includegraphics[scale=0.4]{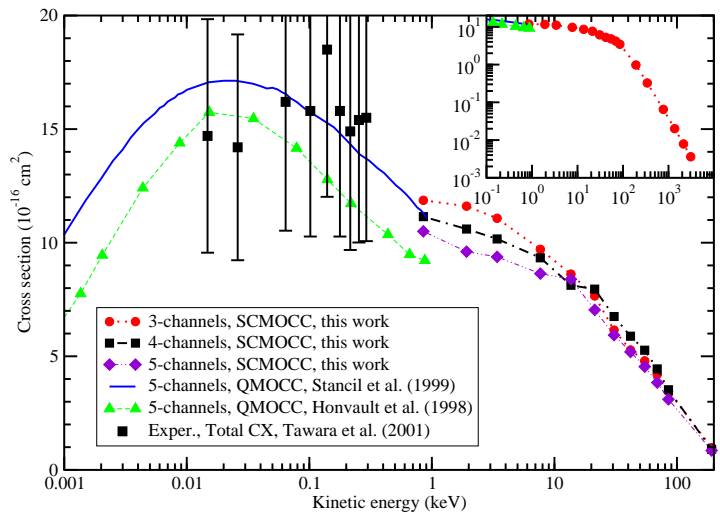}
\caption{The Si$^{3+}$ + He charge exchange cross section for the
$2\rightarrow 1$ transition comparing the current SCMOCC results to
earlier QMOCC results. Note the cross section is given as a function
of center-of-mass kinetic energy and the results of Ref.~\cite{sta99} used
the same diabatic potential as the current work. The experiment is
for total charge exchange.}
\label{cross}
\end{center}
\end{figure}

Figure~\ref{cross} plots the cross section, obtained by integrating the
probability distributions from Figure~\ref{pbsihe3}, for capture to Si$^{2+}$($3s^2~^1S$) (the
2$\rightarrow$1 transition) which dominates the total charge exchange as
the 2$\rightarrow$3 and 2$\rightarrow$4 transitions give small cross sections.
The current calculations using the SCMOCC approach are in
very good agreement with our earlier QMOCC calculation and a computation
performed by Houvault et al. \cite{hon98}. A similar
scattering method was adopted in Ref.~\cite{hon98}, but with different diabatic potentials. The ion beam - gas cell
measurement of Tawara et al \cite{taw01} is consistent with all of the calculations, though
the uncertainty is rather large. Figure~\ref{cross} also illustrates a channel convergence
study where the integral cross section appears to be approaching convergence by $n=5$,
but additional investigations are needed to confirm this result. 

There is a second measurement, but of the rate coefficient at 3900 K in an ion trap
\cite{fan97}. A rate coefficient is obtained by averaging the cross section over
a Maxwellian velocity distribution. 3900 K corresponds to a center-of-mass
kinetic energy of about 0.4 eV and therefore too low of an energy for our SCMOCC
method to be valid. However, as Fig.~6 of Ref.~\cite{sta99} shows, the QMOCC results
are consistent with the ion trap measurement suggesting that the adopted potentials
are reliable. Note also that there is considerable uncertainty in the ion trap
temperature.

The electron capture cross sections to the Si$^{2+}$($3s3p~^3P^o$) (the
2$\rightarrow$3 transition)  are given in the Supplement. The
magnitude of the cross sections are about a factor of 20 smaller than
for the 2$\rightarrow$1 transition and no experimental data exists. There is
reasonable agreement between all calculations, but additional
channels are typically required to
get small cross sections converged. In summary, the Si$^{3+}$ + He charge exchange
system can be an important test case for application to SES devices with $n=3-5$.

\subsection{O$^{7+}$ + H}

\begin{figure}[h]
\begin{center}
\includegraphics[scale=0.40]{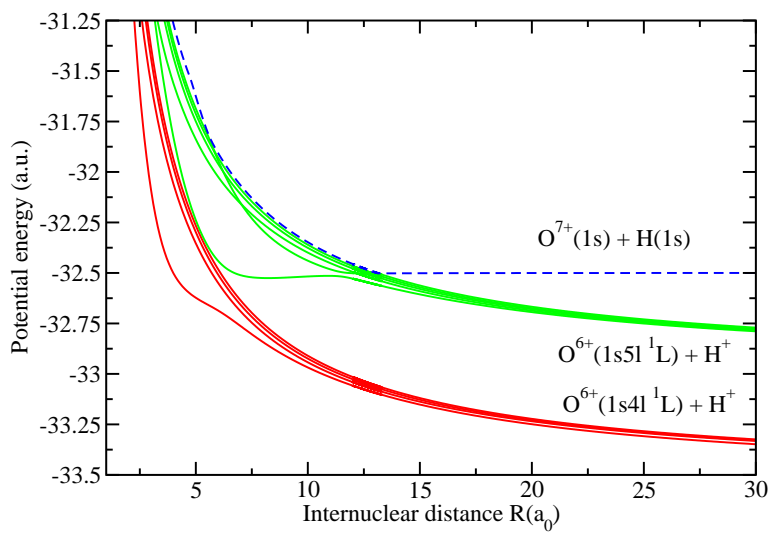}
\caption{Adiabatic potentials for the OH$^{7+}$ singlet system from Nolte et al \cite{nol16}.
The initial channel (10) is given by the blue dashed curve.
}
\label{poto7h}
\end{center}
\end{figure}

Moving to a somewhat larger system, we consider the charge exchange
interaction
\begin{eqnarray}
{\rm O}^{7+}(1s~^2S) + {\rm H}(1s~^2S) & \rightarrow &  {\rm O}^{6+}
(1s5\ell~^1L) + {\rm H}^+,\\
& \rightarrow &  {\rm O}^{6+}
(1s4\ell~^1L) + {\rm H}^+,
\end{eqnarray}
which was studied with the QMOCC method by Nolte et al. \cite{nol16}.
Here we consider the singlet spin system with nearly degenerate principal
quantum number manifolds of 4 and 5 states, giving a total of $n=10$ channels.
The singlet adiabatic potential energies are given in Figure~\ref{poto7h}, while
the full diabatic potential matrix is available from Ref.~\cite{nol16}.

\begin{figure}[h]
\begin{center}
\includegraphics[scale=0.40]{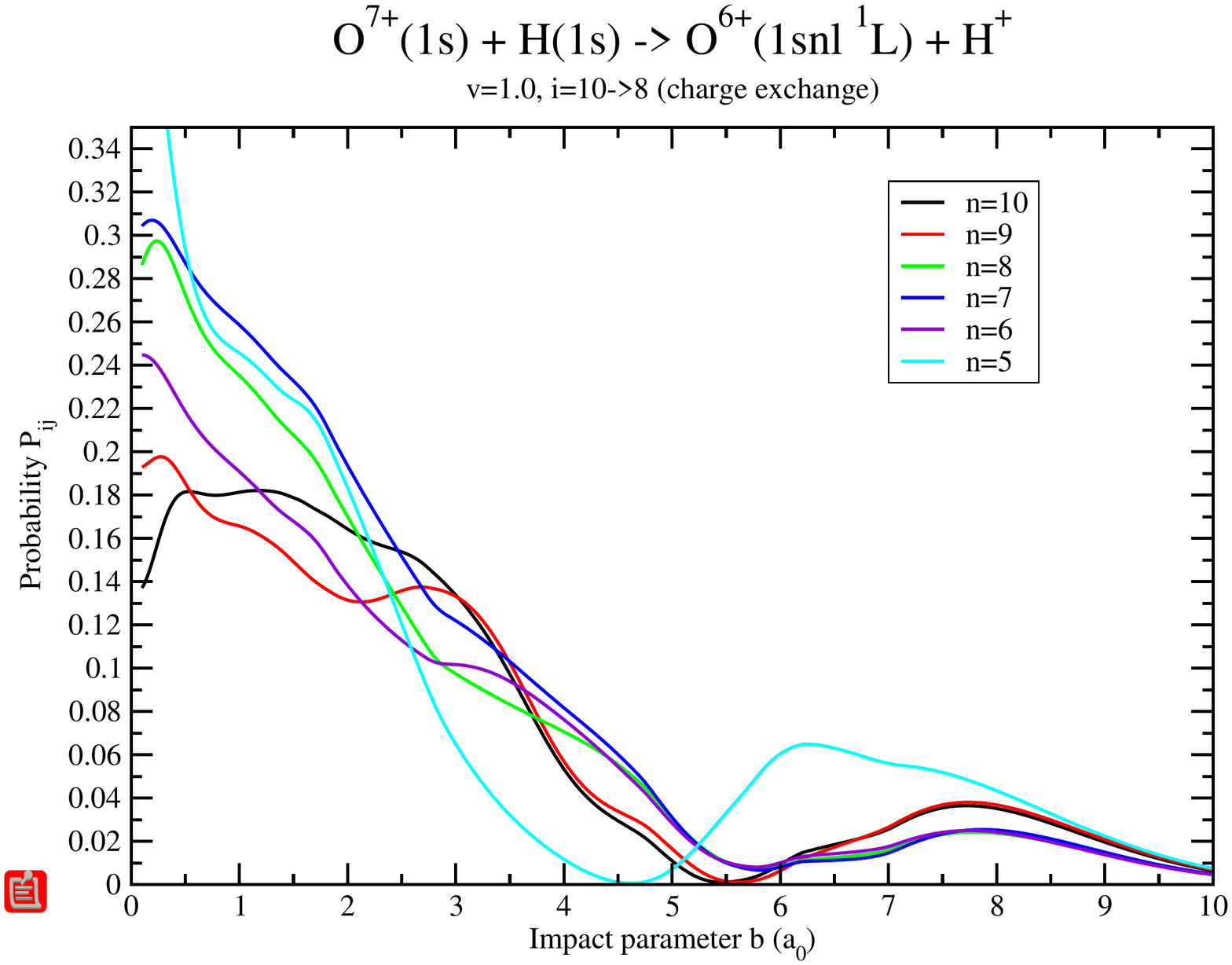}
\caption{The O$^{7+}$ + H probabilities versus impact parameter for the state-resolved charge
exchange reaction with product O$^{6+}$($1s5d~^1D$) + H$^+$ obtained from SCMOCC
calculations
with $n=5-10$ channels and $v_0=1.0$ a.u.}
\label{oh71}
\end{center}
\end{figure}

A series of $n=5$ to $n=10$ channel calculations were performed for $v_0=1.0$ a.u.
Figure~\ref{oh71} displays the charge exchange probability for the $10\rightarrow 8$
transition whose final state is O$^{6+}$($1s5d~^1D$) + H$^+$.  A similar plot
for elastic scattering is shown in Figure~\ref{oh73} with additional results given
in the Supplement. 

There is considerable variation in the probabilities with basis
size so that this collision  system would serve as an interesting test bed for
SES devices of moderate size from $n=5-10$. Further, as the size of  the
system is increased from Na+He to Si$^{3+}$ + He to O$^{7+}$ + H, the
maximum internal energy difference increases with values of 2.1, 8.8, and
28.5 eV, respectively, allowing for about an order of magnitude in range of energy
 scale mapping to the SES (see below).

\begin{figure}[h]
\begin{center}
\includegraphics[scale=0.40]{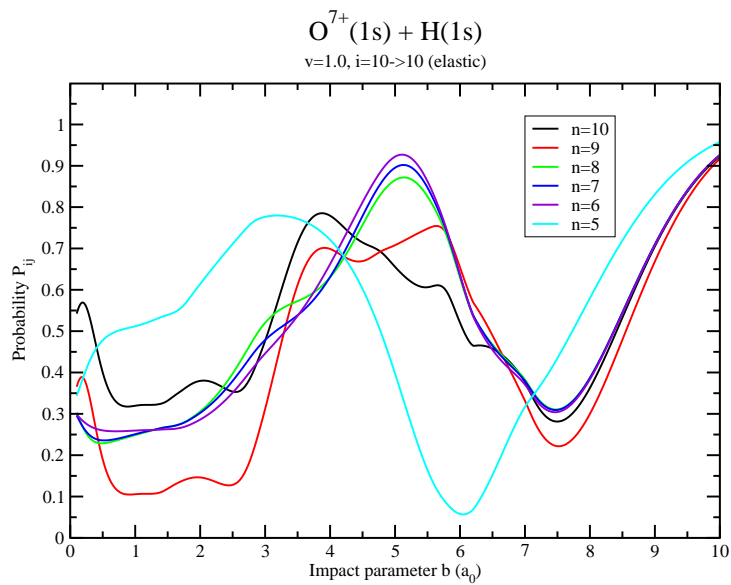}
\caption{The O$^{7+}$ + H probabilities versus impact parameter for elastic
scattering obtained from SCMOCC calculations
with $n=5-10$ channels and $v_0=1.0$ a.u. Note the elastic probability should go to 1.0
as $b\rightarrow \infty$ and is not converged for $b=10$ a$_0$.}
\label{oh73}
\end{center}
\end{figure}

\subsection{Low Kinetic Energies: Improvements to the Classical Simulation}

\begin{figure}[htb]
\begin{center}
\includegraphics[scale=0.4]{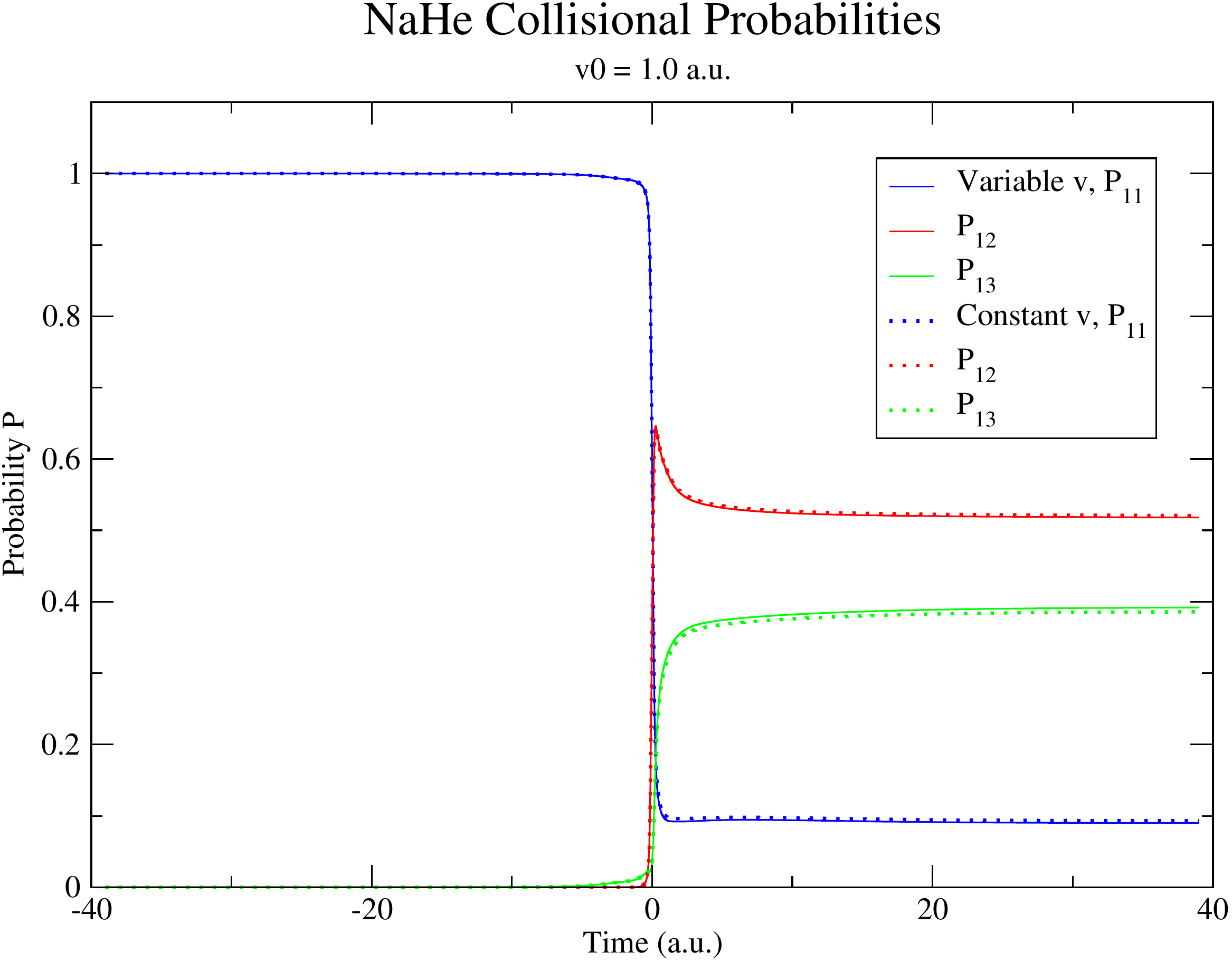}
\caption{Comparison of Na+He scattering probability evolution with collisions time for constant and variable velocities. 
Results are presented for $v_0=1.0$ a.u., $b=1.0$~a$_0$, and $n=3$.}
\label{vint2}
\end{center}
\end{figure}

\begin{figure}[htb]
\begin{center}
\includegraphics[scale=0.4]{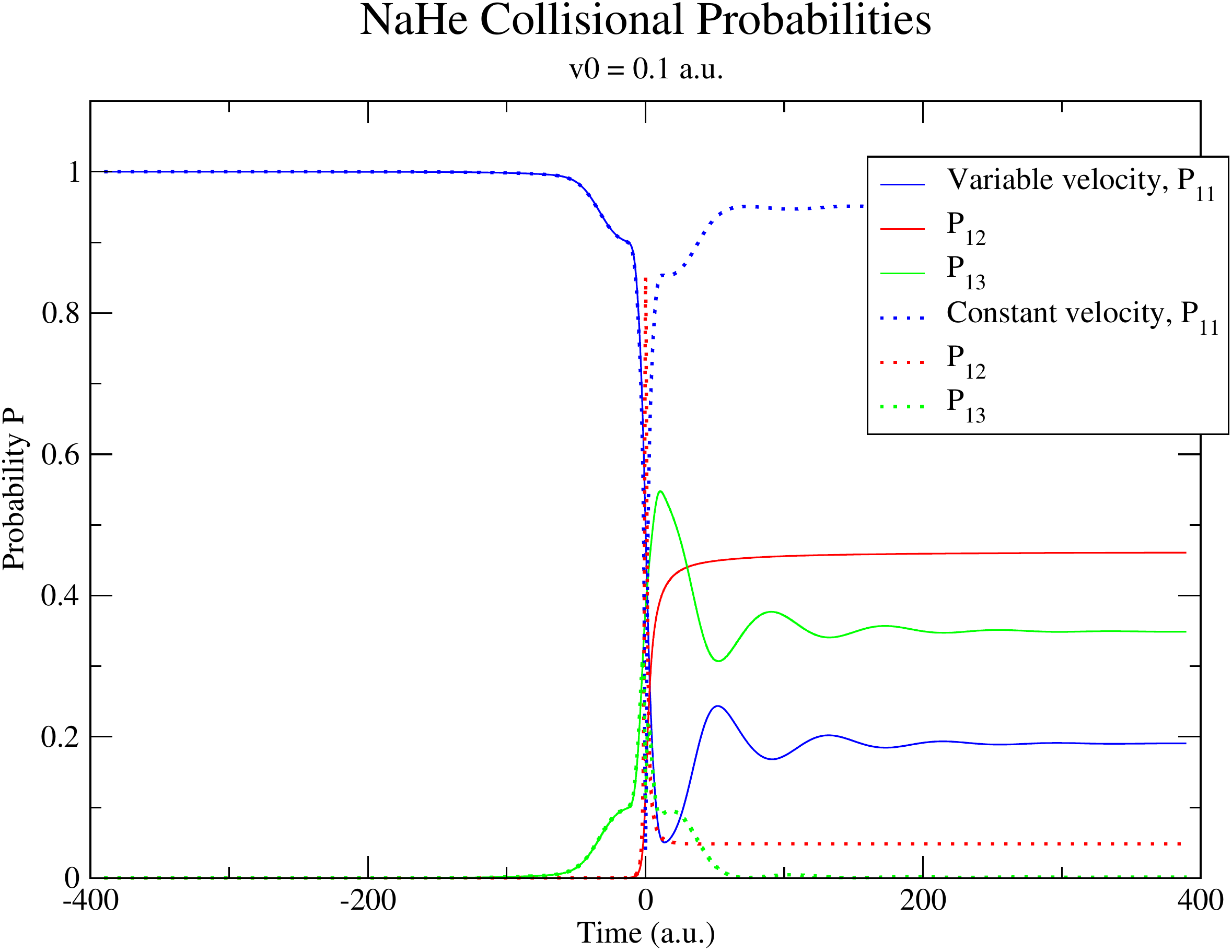}
\caption{Same as Figure~\ref{vint2} but for $v_0=0.1$ a.u.}
\label{vint1}
\end{center}
\end{figure}

\begin{figure}[htb]
\begin{center}
\includegraphics[scale=0.4]{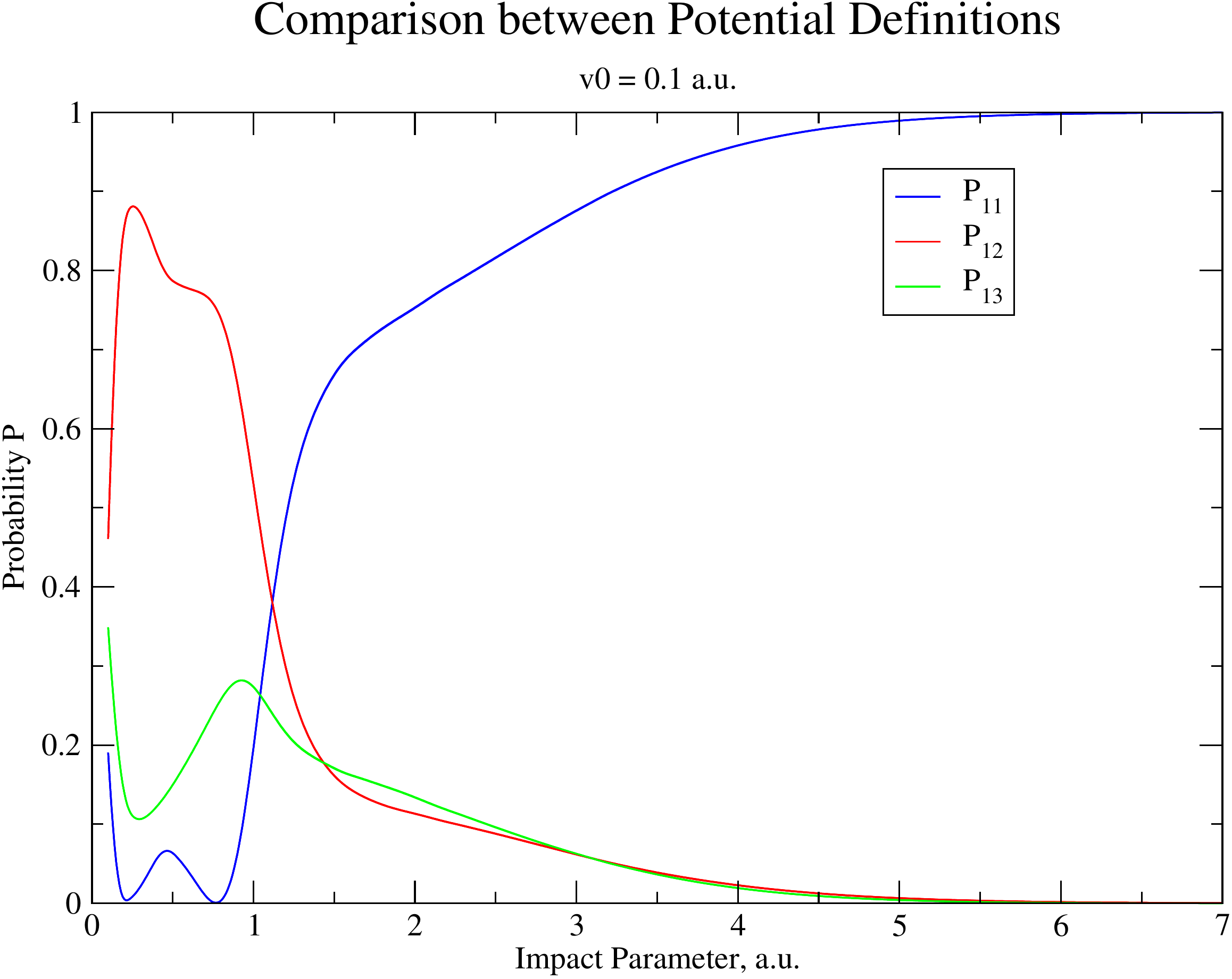}
\caption{Comparison Na+He scattering probabilities versus impact parameter 
for a variable velocity with arithmetically-averaged potentials. Results are presented for $v_0=0.1$ a.u. and
$n=3$.}
\label{potentialavg1}
\end{center}
\end{figure}

The above SCMOCC method assumed a straight line, constant velocity trajectory.  This approach is
expected to break-down for kinetic energies between 0.1 and 1 keV for the above considered
collision systems involving electronic transitions. Or in other words, the straight line SCMOCC method
is probably valid for kinetic energies an order of magnitude larger than the maximum internal energy
difference
\begin{equation}
K \gtrsim 10 |E_f-E_i| = 10|\Delta E|.
\end{equation}
To extend the reliability of the SCMOCC method to lower energies,  curvilinear trajectories can be adopted by solving 
for the velocity $dR/dt$ via 
\begin{equation}
\frac{dR}{dt}= \pm v_0 \sqrt{1 - \frac{b^2}{R^2} - \frac{V(R)}{E}}
\label{velocityint}
\end{equation}
at each step of the time integration. Here, $v_0$ is the initial velocity at infinity, and $E$ is the total energy of the system. However, solving the additional equation~(\ref{velocityint}) increases the computational time, particularly for small values of $v_0$. This is related to the requirement of additional time steps in order to stabilize the final probabilities. Figure~\ref{vint2} displays the probability evolution for the Na+He system as a function of collision time comparing the constant velocity case at $v_0$=1.0 a.u. to use of equation~(\ref{velocityint}), both for $b=1.0$~a$_0$.
At this high velocity, the probabilities are almost identical as expected, but lowering $v_0$ to 0.1 a.u. as shown
in Figure~\ref{vint1}, results in significant differences. The probabilities versus impact parameter for given
in Figure~\ref{potentialavg1}.

\begin{figure}[htb]
\begin{center}
\includegraphics[scale=0.4]{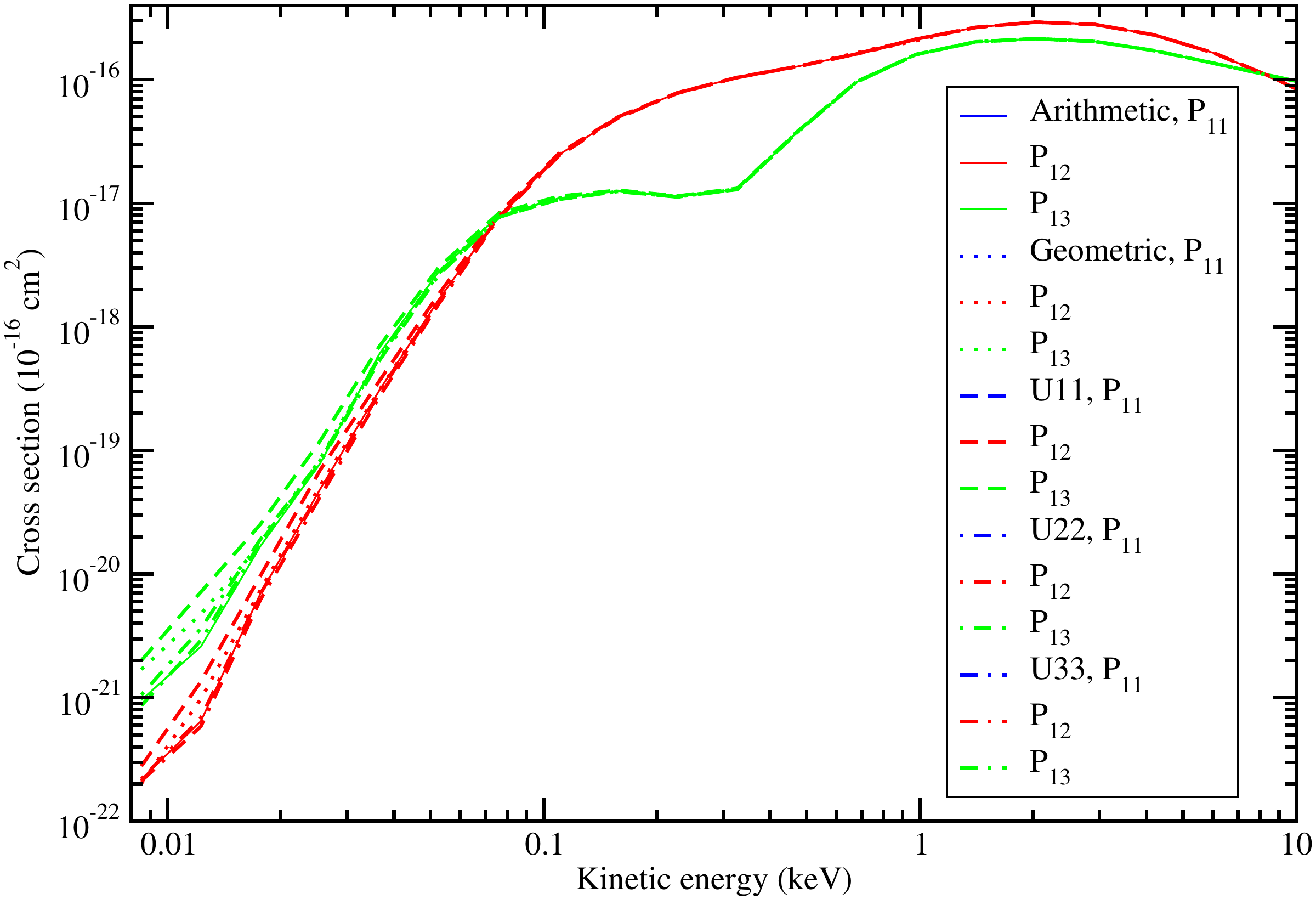}
\caption{Comparison of the $n=3$ Na+He total cross sections for various potential
averaging methods i) arithmetic, ii) geometric, iii) using channel $i=1$, iv) using $i=2$, and v) using $i=5$.
Note elastic cross sections are not shown as the various methods give identical results.} 
\label{potentialavg2}
\end{center}
\end{figure}

While for single channel calculations (i.e., $n=1$) there is no ambiguity in equation~(\ref{velocityint}),
 for multichannel cases,
the $V(R)$ term has presented a theoretical dilemma for many decades. $V(R)$ must be replaced by some superposition,
$\bar{V}(R)$, over all  channel diagonal diabatic potentials, but  the exact prescription is unknown. A number of
authors \cite{bat70, del72, ril73, lia75} have proposed various schemes including:  (i) an arithmetic average, (ii) a geometric average, or (iii) setting $\bar{V}(R)$ equal to the $V_{ii}(R)$ of an individual channel. We find that
the probabilities and cross sections computed by any of these schemes are practically indistinguishable down to a
kinetic energy of $\sim$0.1 keV.  In fact, Delos et al
\cite{del72} suggested that the exact averaging prescription is unimportant as long as some type of averaging is
taken into account, as illustrated in Figures~\ref{vint2}-\ref{potentialavg1}. 
For energies less than $\sim$0.1 keV, Figure~\ref{potentialavg2} shows some dispersion of the cross sections
for different averaging schemes, but in this energy regime the cross sections become very small. Nevertheless, the arithmetic and geometric approaches give similar results and therefore we
default to an arithmetic average of all channels for $\bar{V}(R)$.

\subsection{Pushing to Even Lower Kinetic Energies: Ehrenfest Symmetrization}

\begin{figure}[h]
\begin{center}
\includegraphics[scale=0.4]{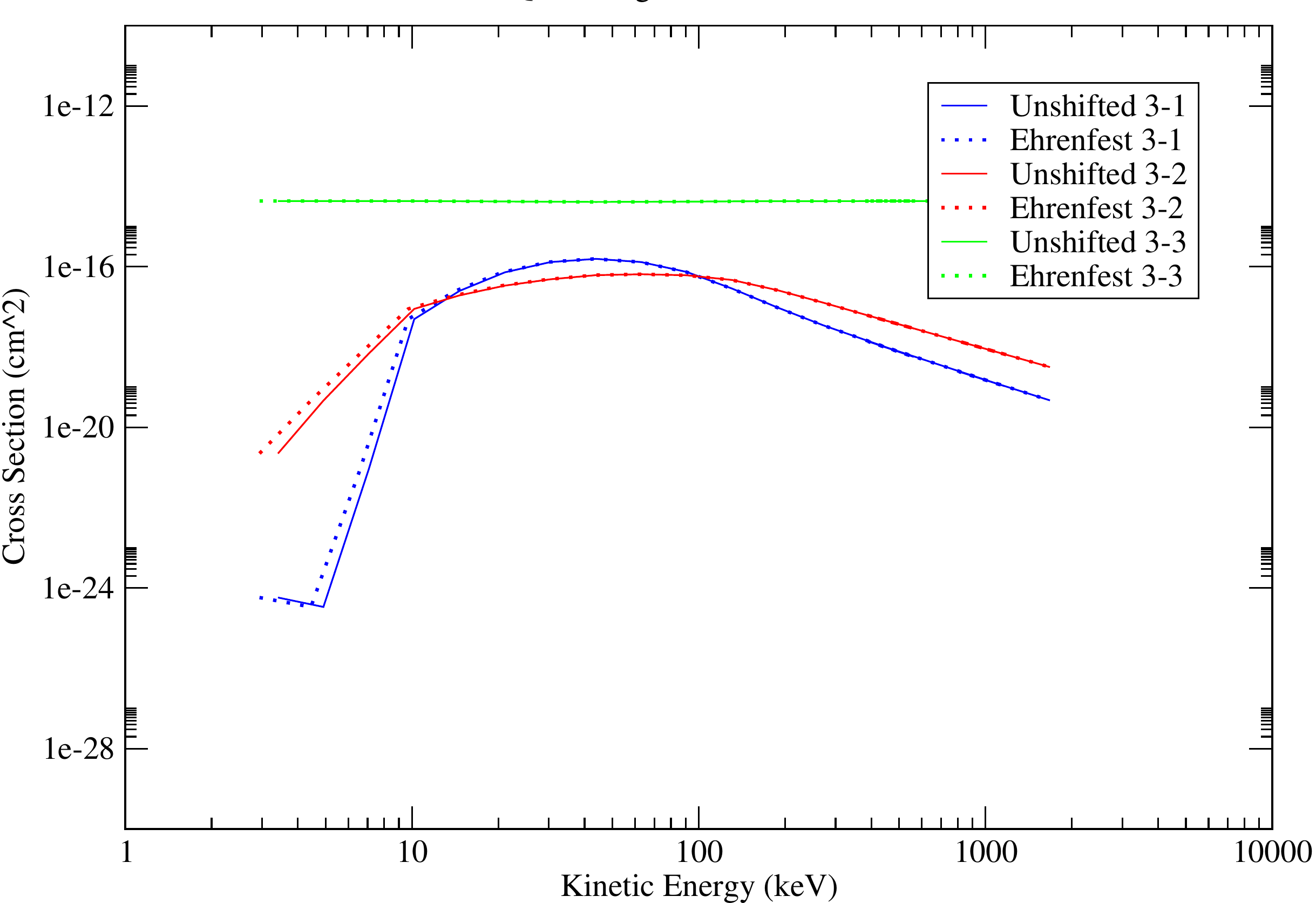}
\caption{Integral cross sections for Si$^{3+}$ + He with variable velocity illustrating the effect of the symmetrized Ehrenfest approach. As expected, the elastic cross sections are unaffected.}
\label{ehrenfest}
\end{center}
\end{figure}

The application of potential averaging is clearly not a robust theoretical procedure and in
fact introduces a problem in which the principle of detailed balance,
\begin{equation}
P_{if}(E) = P_{fi}(E),
\end{equation}
may be violated \cite{bill03}. This deficiency becomes worse with decreasing collision energy
thereby limiting the applicability of the SCMOCC method. Billing \cite{bill03} has proposed a
correction by introducing a so-called symmetrized Ehrenfest approach. The method, which
is not related to Ehrenfest's Theorems,
shifts the relative velocity for a given initial channel and has been shown to give reliable cross
sections for a variety of collision systems \cite{bill03,sem13,iva14,bab16}. We apply it here
as a post-processing algorithm which redefines the kinetic energy $K$ of the collision system. For a 2-state case, the kinetic energy is redefined according to 
\begin{equation}
E = \bar{K}+\frac{\Delta E}{2}+\frac{\Delta E^2}{16\bar{K}},
\label{ehreneq}
\end{equation}
where $\bar{K}$ is the redefined kinetic energy. It is presumed to be valid for $K\ge \Delta E /4$. So, that for an endoergic process,
$E = \Delta E$ with $\Delta E > 0$, while $E=0$ and $\Delta E < 0$ for exoergic transitions. As an example,
Figure~\ref{ehrenfest} compares the integral cross sections for an $n=3$ computation of Si$^{3+}$ + He
with and without the symmetrized Ehrenfest approximation for the case of arithmetic-averaged potentials.
The former is larger than the uncorrected case for inelastic transitions below 10 keV with the difference increasing
with decreasing $K$. In summary, explicitly solving for the relative velocity as a function of time with arithmetic-averaged multichannel potentials and a post-processing shift of the kinetic energy via the symmetrized
Ehrenfest approach should allow the SCMOCC method to be applied to compute cross sections for
kinetic energies a factor $\sim$100 smaller than with the straight line, constant velocity approximation,
that is, approaching kinetic energies as low as 1 eV.

\section{Molecular Collisions on an SES Processor}

Now we turn to simulating collision problems on a quantum computer using the
SES approach.

\subsection{Hamiltonian Mapping}

To illustrate the SES simulation procedure, we consider the $n=5$ Si$^{3+}$+He charge exchange collision process~(\ref{sihereaction}).  The collision Hamiltonian $h(t)$ must be rescaled by energy using the method 
described in Geller et al \cite{Geller2015} so that the rescaled Hamiltonian  ${\cal H}(t)$ is compatible with the
SES processor, 
\begin{equation}
{\cal H}(t)=\frac{h(t)-c(t)\times I}{\lambda (t)},
\end{equation}
where $c(t)$ is the mean of diagonal elements of $h(t)$  and $\lambda(t)$ is the rescaling function such that each matrix element of the SES Hamiltonian ${\cal H}(t)$ lies
within the characteristic energy range of the SES device. The simulated (quantum computer) time $t_{qc}$ on the SES processor satisfies a nonlinear relation with respect to the physical time $t$,
\begin{equation}
t_{qc}(t)=\int_{0}^{t} \lambda dt^{\prime}.
\end{equation}

\begin{figure}[htb]
\begin{center}
\includegraphics[scale=0.3]{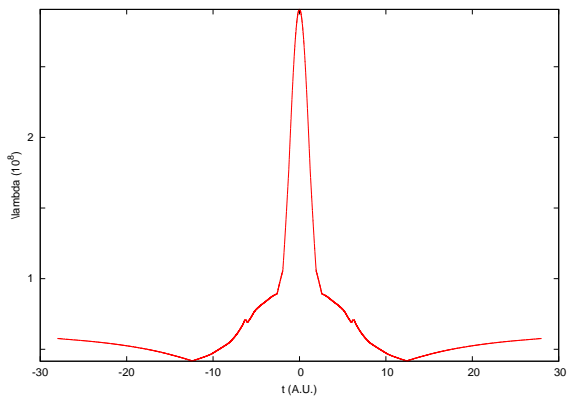}
\caption{Rescaling factor $\lambda$ as a function of the physical time $t$ for the $n=5$ Si$^{3+}$+He charge exchange collision simulation. Collision parameters are chosen as $b=1.0$ a$_0$ and $v_0=0.5$ a.u.}
\label{t_vs_lambda}
\end{center}
\end{figure}

\begin{figure}[htb]
\begin{center}
\includegraphics[scale=0.35]{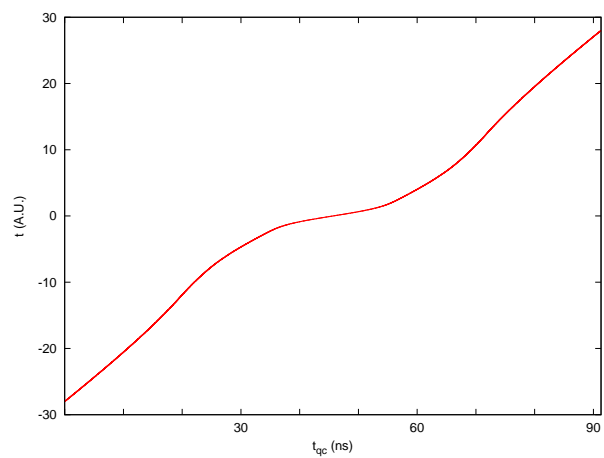}
\caption{The simulated time $t_{qc}$ as a nonlinear function of the physical time $t$ for the $n=5$ Si$^{3+}$+He charge exchange collision simulation. Collision parameters are chosen as  $b=1.0$ a$_0$ and $v_0=0.5$ a.u.}
\label{tqc_vs_tAtom}
\end{center}
\end{figure}

\begin{figure}[h]
\begin{center}
\includegraphics[scale=0.4]{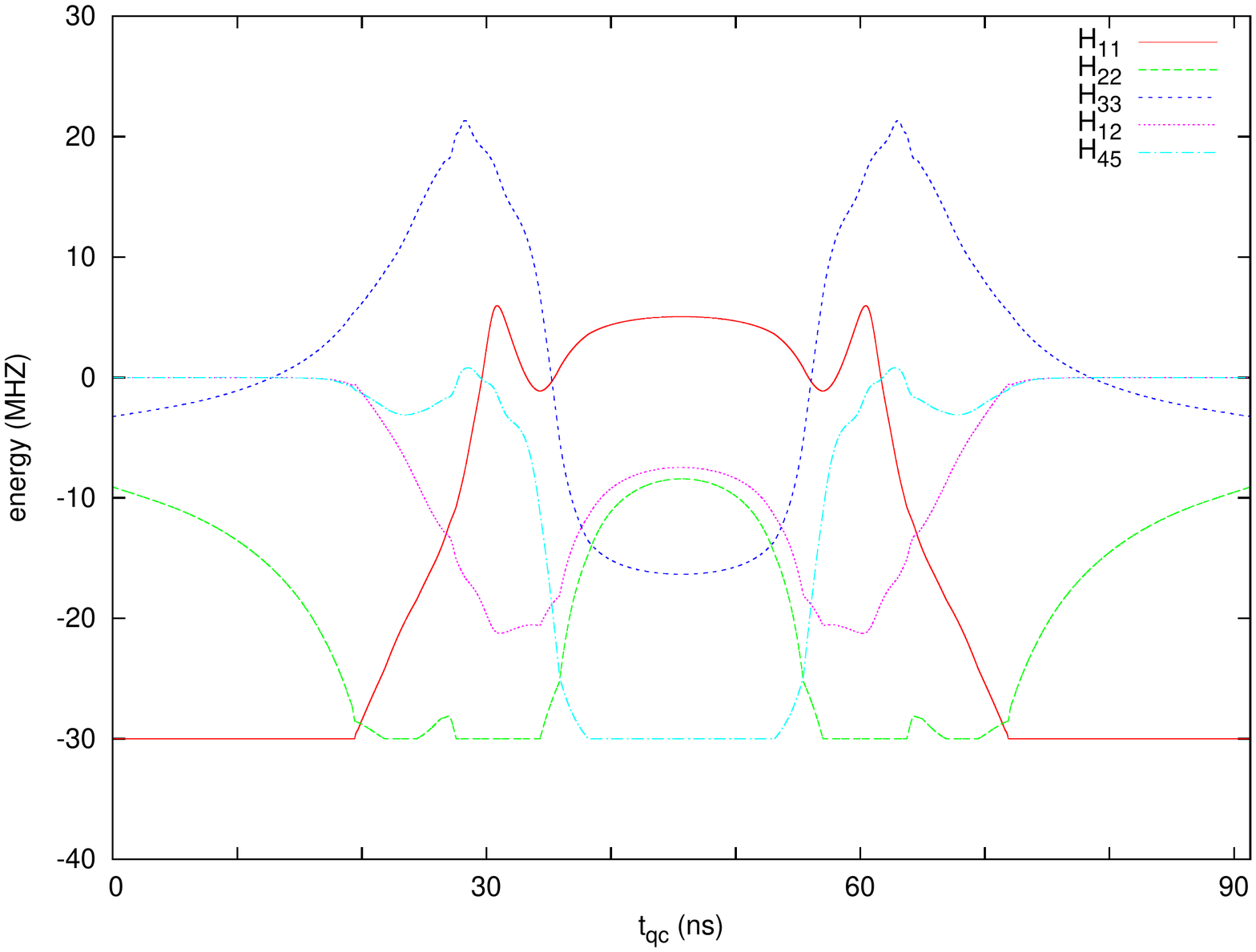}
\caption{Matrix elements of the SES Hamiltonian $\cal{H}$ as a function of the simulated time $t_{qc}$ for the 
Si$^{3+}$+He charge exchange collision simulation. Collision parameters are chosen as $b=1.0$ a$_0$ and $v_0=0.5$ a.u.}
\label{SES_HvsTqc}
\end{center}
\end{figure}

The rescaling function $\lambda$ is shown in Figure~\ref{t_vs_lambda} , and the nonlinear time relationship is shown in Figure~\ref{tqc_vs_tAtom}. As can be seen from Figure~\ref{t_vs_lambda}, near the peak of the rescaling function, the collision energy scale is large where the dynamics becomes significant. With use of the rescaling function, the SES Hamiltonian matrix elements are
obtained as shown in Figure~\ref{SES_HvsTqc}.

\subsection{Scattering Algorithm and Simulation}

\begin{figure}[htb]
\begin{center}
\includegraphics[scale=0.4]{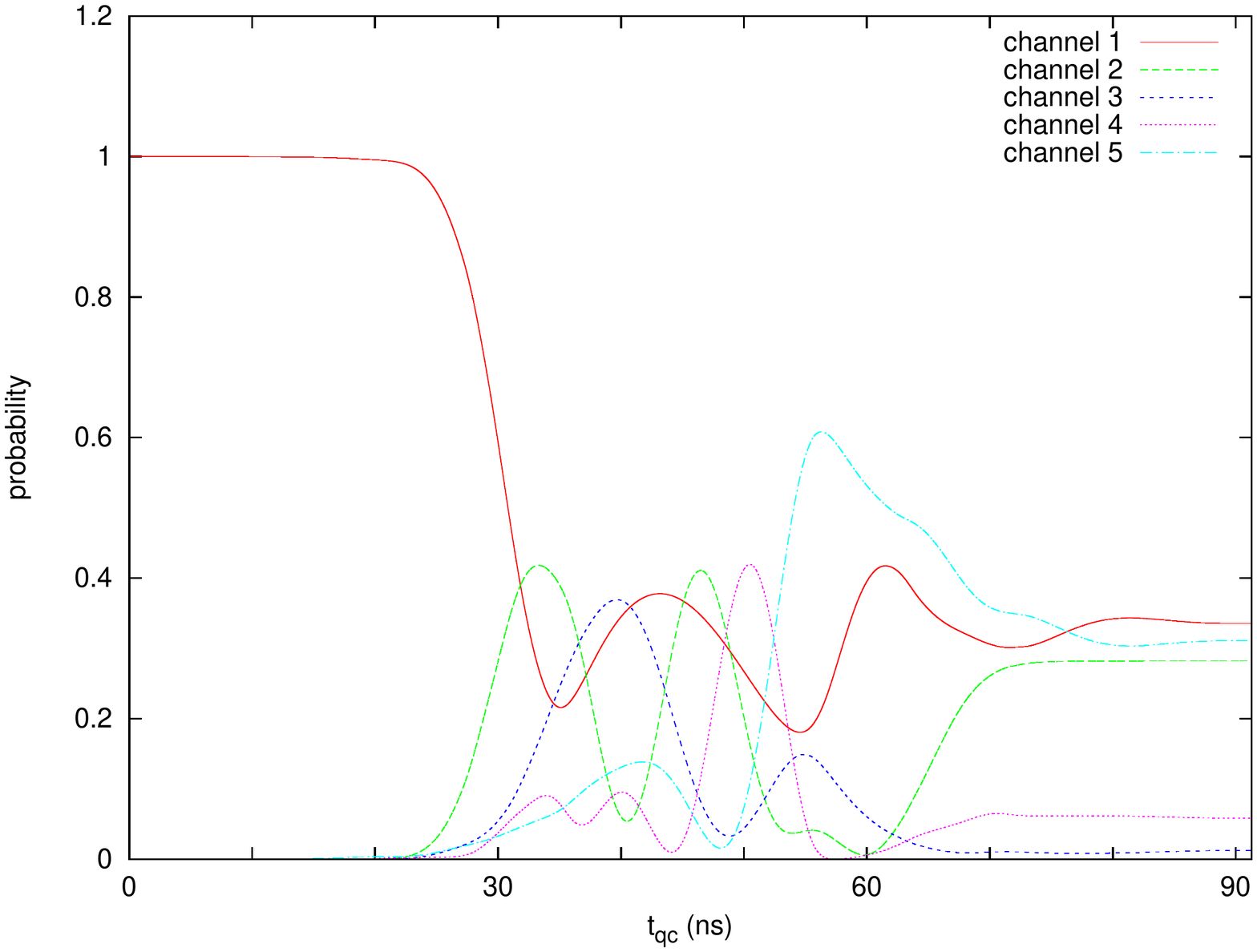}
\caption{Scattering probabilities versus simulation time on a SES processor for a $n=5$ Si$^{3+}$+He collision with  $b=1.0$ a$_0$ and $v_0=0.5$ a.u.}
\label{tqc_vs_prob}
\end{center}
\end{figure}

Mapping the collision Hamiltonian to the SES Hamiltonian $\cal{H}$ by use of the rescaling function given in  Figure~\ref{t_vs_lambda} and its implementation in a SES processor results in scattering probabilities. Figure~\ref{tqc_vs_prob} depicts the probabilities computed on a classical computer for a simulation of the SES processor for $n=5$. Compared with Figure~\ref{tAtom_vs_prob}, we see that the dynamics near the time of closest approach are rescaled on the SES processor and occupy most of the simulation.  Table 1 shows the final probabilities for transitions out of state 1 on a classical computer and for a simulation of the SES method. These results indicate that the accuracy of the SES method is comparable with that of the classical simulation and the
relative error increases with decreasing collision probability.
One should note that by mapping the Si$^{3+}$+He collision problem to the SES processor, a single run of the simulation can be completed in about only $70$ ns, independent of $n$. 

\begin{table}[!hbp]
\begin{center}
    \begin{tabular}{ | l | l | l | p{4cm} |}
    \hline
    Probability & Classical Simulation & SES simulation & Relative Error \\ \hline
    $P_{11}$ & $0.3374 $ & $0.3371$ & 0.095\% \\ \hline
    $P_{12}$ & $0.2784 $ & $0.2813$ & 1.062\% \\ \hline
    $P_{13}$ & $1.2848 \times 10^{-2}$ & $1.3410 \times 10^{-2}$ & 4.369\% \\ \hline
    $P_{14}$ & $5.8881 \times 10^{-2}$  & $5.7782 \times 10^{-2}$ & 1.866\% \\ \hline
    $P_{15}$ & $0.3125$  & $0.3104$ & 0.671\%\\
    \hline
    \end{tabular}
    \caption {Probabilities $P_{if}$ in the $f$-th channel ($i,f=1-5$) for a Si$^{3+}$+He collision in the classical simulator and the SES simulator, respectively. The parameters are chosen as $v_{0}=0.5$ a.u. and $b=1.0$ a$_0$ with $i=1$ the initial channel.}
\end{center}
\end{table}

\section{Conclusions}

As a potential application for quantum simulation using the single excitation subspace (SES)
approach, molecular collisions involving two-atom systems with increasing Hamiltonian
dimension are studied using a standard semiclassical
molecular-orbital close-coupling (SCMOCC) scattering method. These systems are first
studied on a classical computer and then simulations of SES processors are performed.
The $n=3$ qubit/molecular channel Na+He excitation problem is extended beyond our earlier work \cite{Geller2015}
to computations of $n=3-5$ for Si$^{3+}$ + He charge exchange and $n=5-10$ O$^{7+}$ + H charge exchange.
Good agreement is found between final probabilities and cross sections from the classical and SES
simulations based on straight line trajectories above $\sim$1 keV. 

To extend the simulations to lower energy, we augment the SCMOCC approach with curvilinear trajectories on
various averaged multichannel potentials. Further, to correct for a violation in detailed balance, we explore use
of a symmetrized Ehrenfest approach which, combined with potential averaging, will allow for the study of
collision systems approaching the chemical regime near 1 eV. As a consequence, the application of the
SCMOCC method for quantum simulation with the SES approach appears promising for collision problems
with 10 channels or more on similarly sized SES devices.

As outlined in the Introduction, quantum-mechanical calculations on classical computers have currently
peaked at the treatment of four- and five-atom systems for time-independent (TI) inelastic and time-dependent
reactive scattering, respectively. In the former case, more than 10,000 channels were required, which is
a record as far as we are aware for such calculations. A dream today is to be able to perform TI inelastic
calculations for the five-atom systems H$_2$O+H$_2$ and NO$_2$+OH which could require $\sim$50,000-500,000
channels on a nine-dimensional potential energy surface. As such calculations can only be envisioned on
the next generation of massively parallel CPU/GPU machines, there may be an opportunity for the
SES/SCMOCC quantum simulation approach to attack these problems if the construction and
operation of large $n$-qubit, fully-connected
quantum computers become feasible.

\newpage

\section{Supplement}

\subsection{Propagator Benchmarking}
One of the most interesting applications of the SES method is that of a general-purpose Schr\"odinger equation solver  for TD Hamiltonians \cite{Geller2015}. The total time required to perform a single run of the quantum simulation is
\begin{equation}
t_{qu}=t_{qc}+t_{meas},
\end{equation}
where $t_{meas}$ is the qubit measurement time which is about $100$ ns. Thus the time for a single run is $\sim 200$ ns
independent of the size of the Hamiltonian matrix $n$.

Though the classical runtime depends on a variety of issues, we can still explore the possibility of speedup by benchmarking the time required to classically simulate a TD Hamiltonian. We studied the classical simulation runtime $\tau_{cl}$ for this problem, comparing four standard numerical algorithms:
(i) Crank-Nicholson integration \cite{Thomas1995}, (ii) the Chebyshev propagator \cite{Kosloff1991},
\begin {equation}
e^{-(iHt/ \hbar)} \approx \sum_{n=0}^{N} a_{n} P_{n}(-iHt/ \hbar),
\end {equation}
(iii) Runge-Kutta (RK) integration \cite{press2007} and (iv) time-slicing combined with matrix diagonalization \cite{Geller2015}. Both the fourth order Runge-Kutta and the preconditioned adaptive step-size Fehlberg-Runge-Kutta method introduced in 
Ref. \cite{Tremblay2004}  are used in the RK methods here.  In the preconditioned approach, a constant preconditioner is applied such that the eigenvalues of the preconditioned Hamiltonian $H_{I}$ are small and thus the RK method converges quickly using the form
\begin {equation}
H_{I}= H-I(E_{\rm max}+E_{\rm min})/2,
\end {equation}
where $E_{\rm max}$ and $E_{\rm min}$ are the maximum and minimum eigenvalues of the Hamiltonian $H$, respectively. Here, we use the Gershgorin Circle Theorem \cite{Golub1996} to
estimate the values of $E_{\rm max}$ and $E_{\rm min}$.
For time-slicing combined with diagonalization of a given $H$, the unitary matrix $V$ of its eigenvectors and the diagonal matrix $D$ of its eigenvalues are computed and then $e^{-iHt}$ is obtained by
\begin {equation}
Ve^{-iDt}V^{\dagger}.
\end {equation}
Table 2 gives the computation times for each method. The relative errors, compared to the results from a standard high-precision Crank-Nicholson integrator, are bounded by $2\%$.
We find that the preconditioned adaptive step-size Fehlberg-Runge-Kutta method  is the fastest approach for the specific problem considered here resulting in a speed-up by better than a factor $\sim$220 compared to the standard Crank-Nicholson propagator.

\begin{table}[!hbp]
\begin{center}
    \begin{tabular}{ | p{2cm} | p{2cm} | p{2cm} | p{2cm} | p{2cm} l}
    \hline
    & Crank-Nicholson & Chebyshev & Runge-Kutta & Matrix Diagonalization  \\ \hline
    without precondition, fixed time step & 193.93s (0.001 a.u. per step )& 198.35s (0.001 a.u. per step)
 &137.61s (0.001 a.u. per step) & 1.32 s (0.2 a.u. per step)\\
    \hline
    with precondition, fixed time step &  121.24s (0.05 a.u. per step) & N.A. & 7.75s (0.1 a.u. per step) & 1.31 s (0.2 a.u. per step) \\ \hline
    with precondition, adaptive stepsize, exact eigenvalues&  N.A. & N.A. & 3.80s (Fehlberg-Runge-Kutta  local error bound $10^{-4}$) & 1.35s (double-step adaptive method with local error bound $10^{-4}$) \\ \hline
    with precondition, adaptive stepsize, eigenvalue estimation & N.A. & N.A. & 0.88s (local error bound $10^{-4}$) & N.A. \\
    \hline
    \end{tabular}
    \caption{Computation time for $n=5$ Si$^{3+}$+He collision problem using various numerical methods. Collision parameters are chosen as $b=1.0$ a$_0$ and $v_{0} =0.5$ a.u.}
\end{center}
\end{table}

\subsection{Additional Classical Scattering Results}

\begin{figure}[h]
\begin{center}
\includegraphics[scale=0.4]
{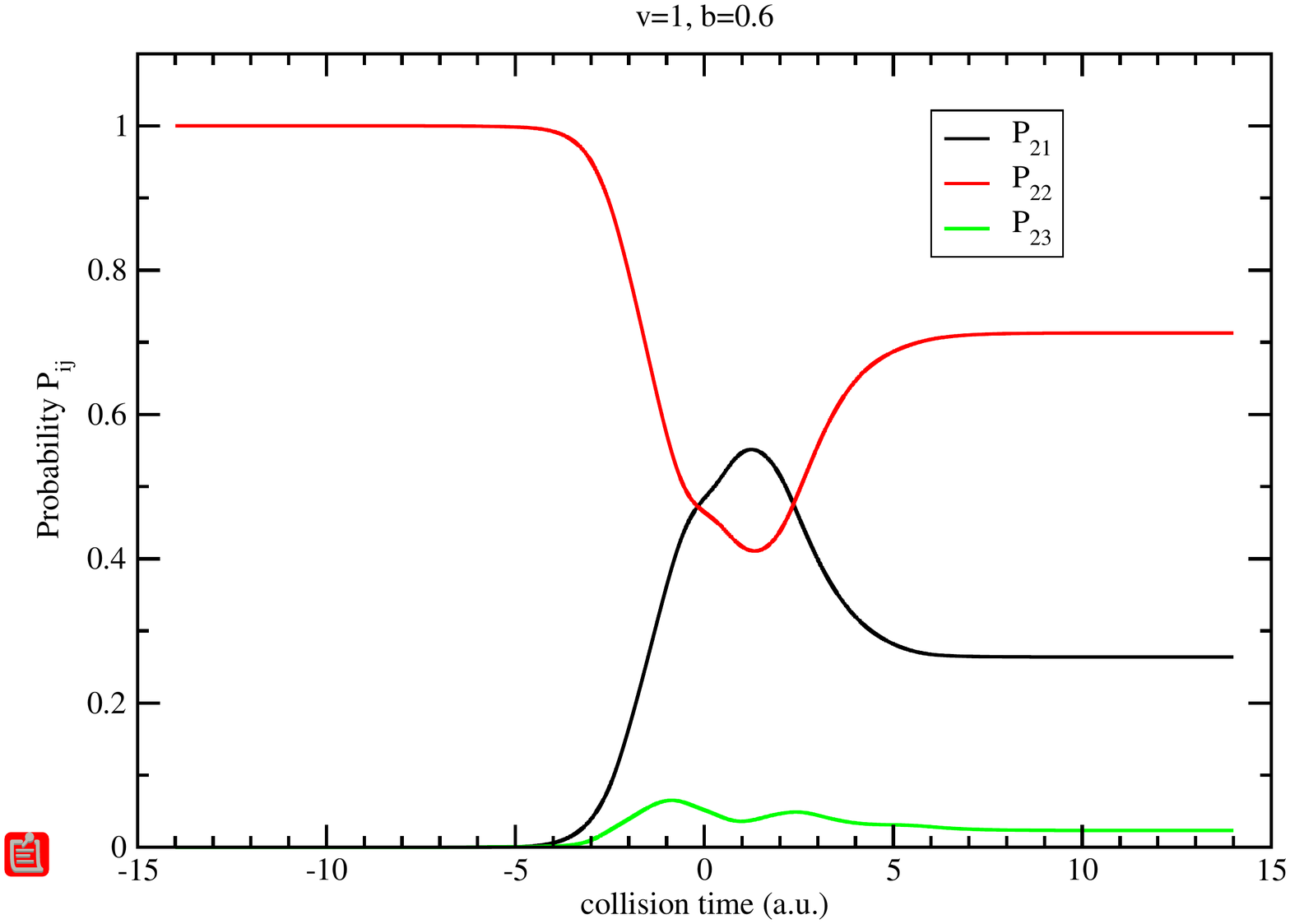}
\caption{The Si$^{3+}$ + He probabilities as a function of collision time for the elastic ($2\rightarrow 2$) and
charge exchange transitions ($2\rightarrow 1$ and $2\rightarrow 3$).
$n=3$ channel case with $b=0.6$ a$_0$, $v_0=1.0$ a.u.}
\label{probsihe3b}
\end{center}
\end{figure}

\begin{figure}[h]
\begin{center}
\includegraphics[scale=0.4]{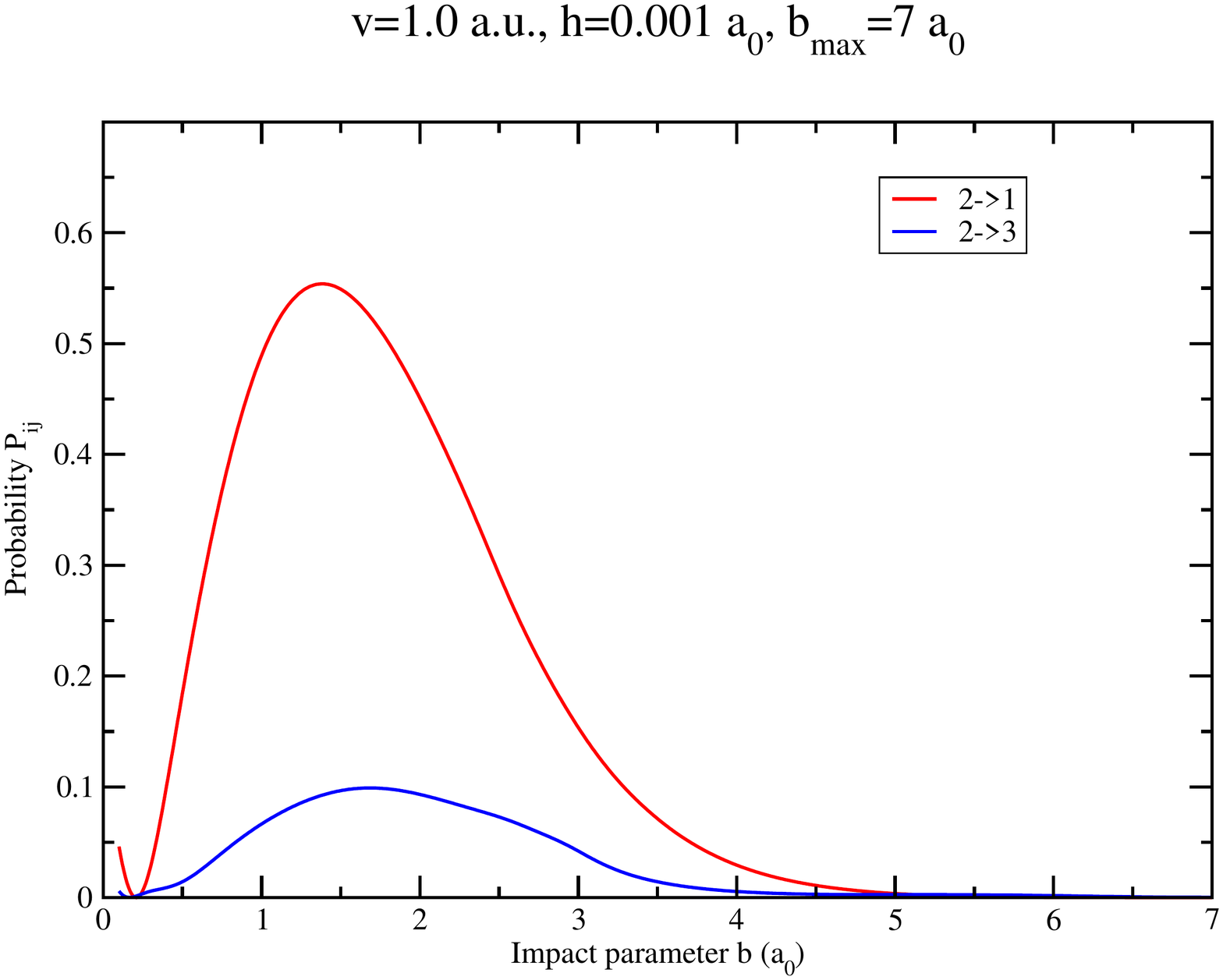}
\caption{The Si$^{3+}$ + He charge exchange probability for the $2\rightarrow 1$ ($2~^2\Sigma^+
\rightarrow 1~^2\Sigma^+$) and $2\rightarrow 3$ ($2~^2\Sigma^+
\rightarrow 3~^2\Sigma^+$) transitions versus
impact parameter. $n=3$ channel case with $v_0=1$ a.u.}
\label{pbsihe3b}
\end{center}
\end{figure}

\begin{figure}[h]
\begin{center}
\includegraphics[scale=0.40]{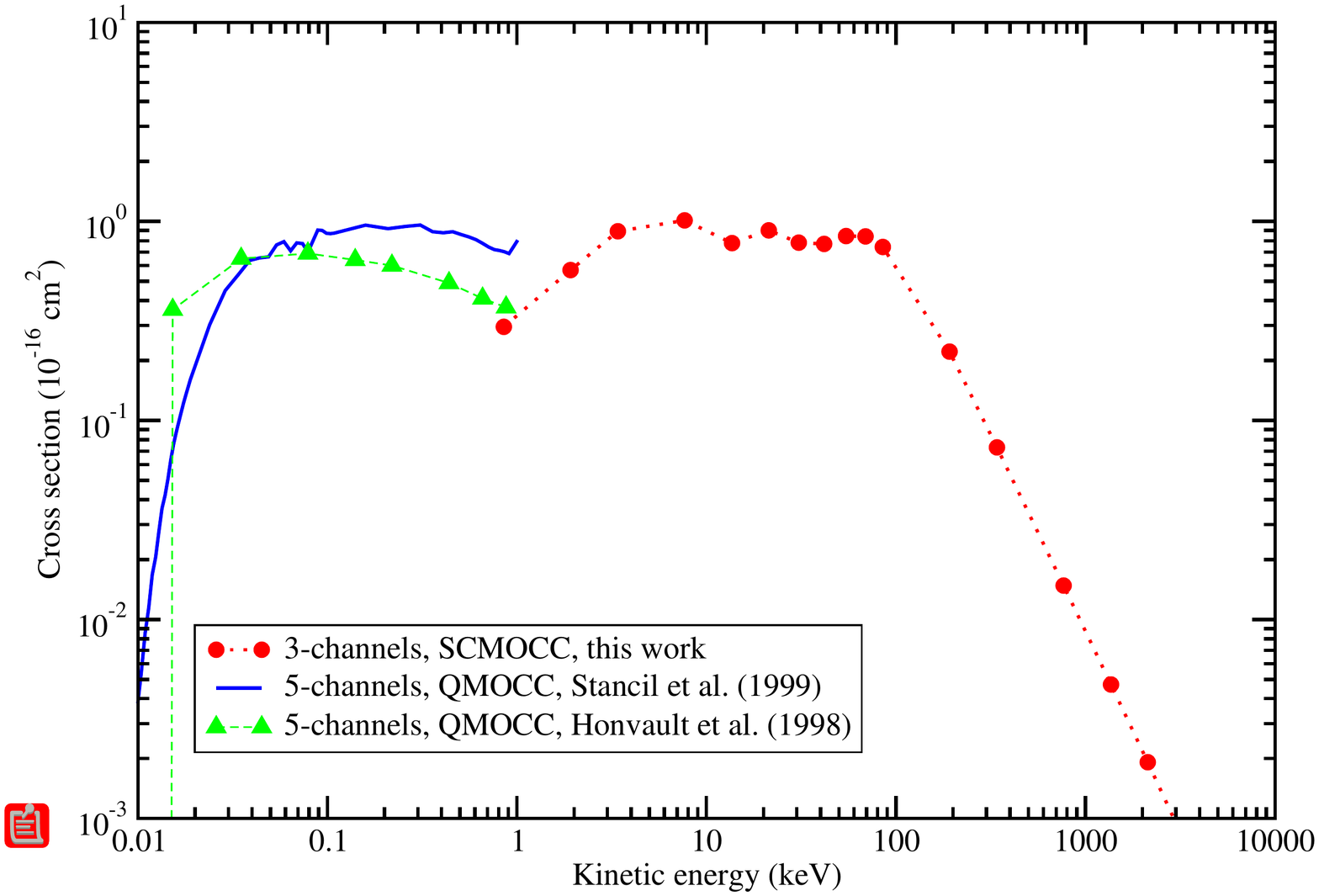}
\caption{The Si$^{3+}$ + He charge exchange cross section for the
$2\rightarrow 3$ transition comparing the current SCMOCC results to
earlier QMOCC results. Note the cross section is given as a function
of center-of-mass kinetic energy and the results of  Ref.~\cite{sta99} used
the same diabatic potentials as the current work.}
\label{cross3P}
\end{center}
\end{figure}

\begin{figure}[h]
\begin{center}
\includegraphics[scale=0.4]{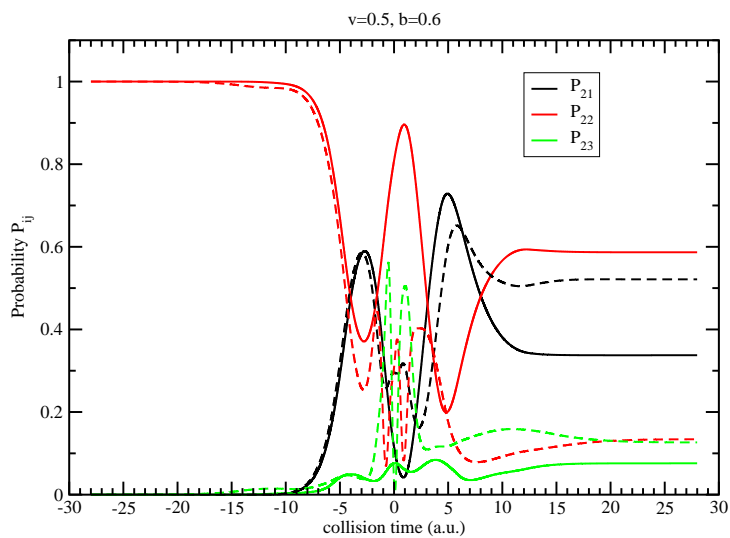}
\caption{Same as Fig.~\ref{probsihe3}, but comparing $n=3$ channels (solid lines) and
$n=4$ channels (dashed lines) for $v_0=0.5$ a.u. and $b=0.6$ a$_0$.}
\label{probsihe3_4chn}
\end{center}
\end{figure}

\begin{figure}[h]
\begin{center}
\includegraphics[scale=0.4]{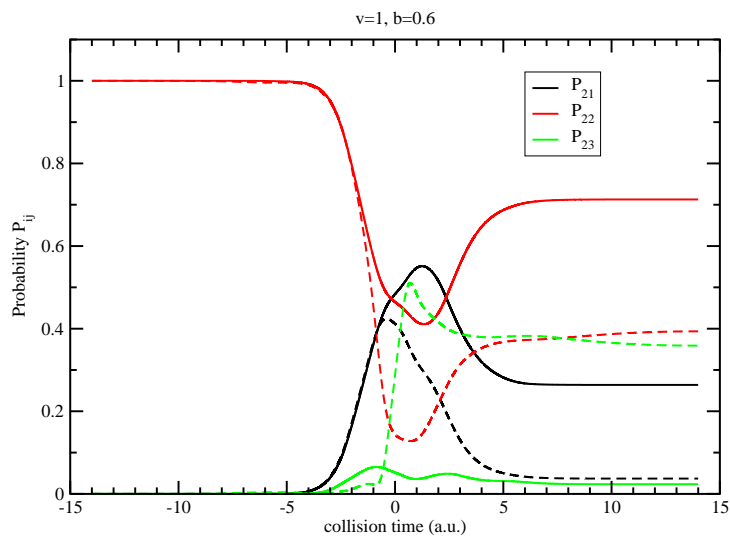}
\caption{Same as Fig. ~\ref{probsihe3_4chn}, but for $v_0=1.0$ a.u. and $b=0.6$ a$_0$.}
\label{probsihe3_4chnb}
\end{center}
\end{figure}

\begin{figure}[h]
\begin{center}
\includegraphics[scale=0.5]{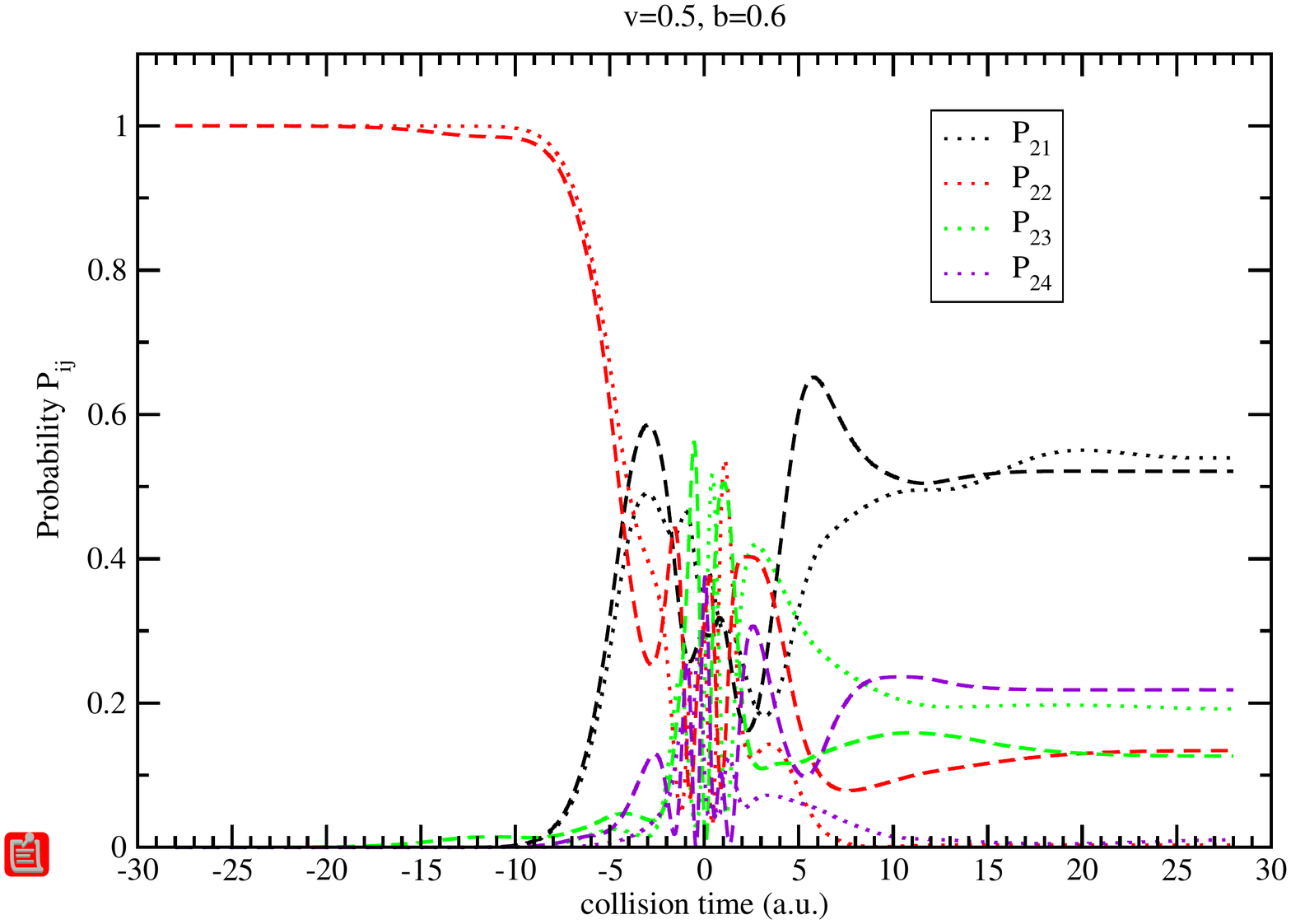}
\caption{Same as Fig.~\ref{probsihe3_4chn}, but comparing $n=4$ channels (dashed lines) and
$n=5$ channels (dotted lines) for $v_0=0.5$ a.u. and $b=0.6$ a$_0$.}
\label{probsihe3_5chn}
\end{center}
\end{figure}

\begin{figure}[h]
\begin{center}
\includegraphics[scale=0.4]{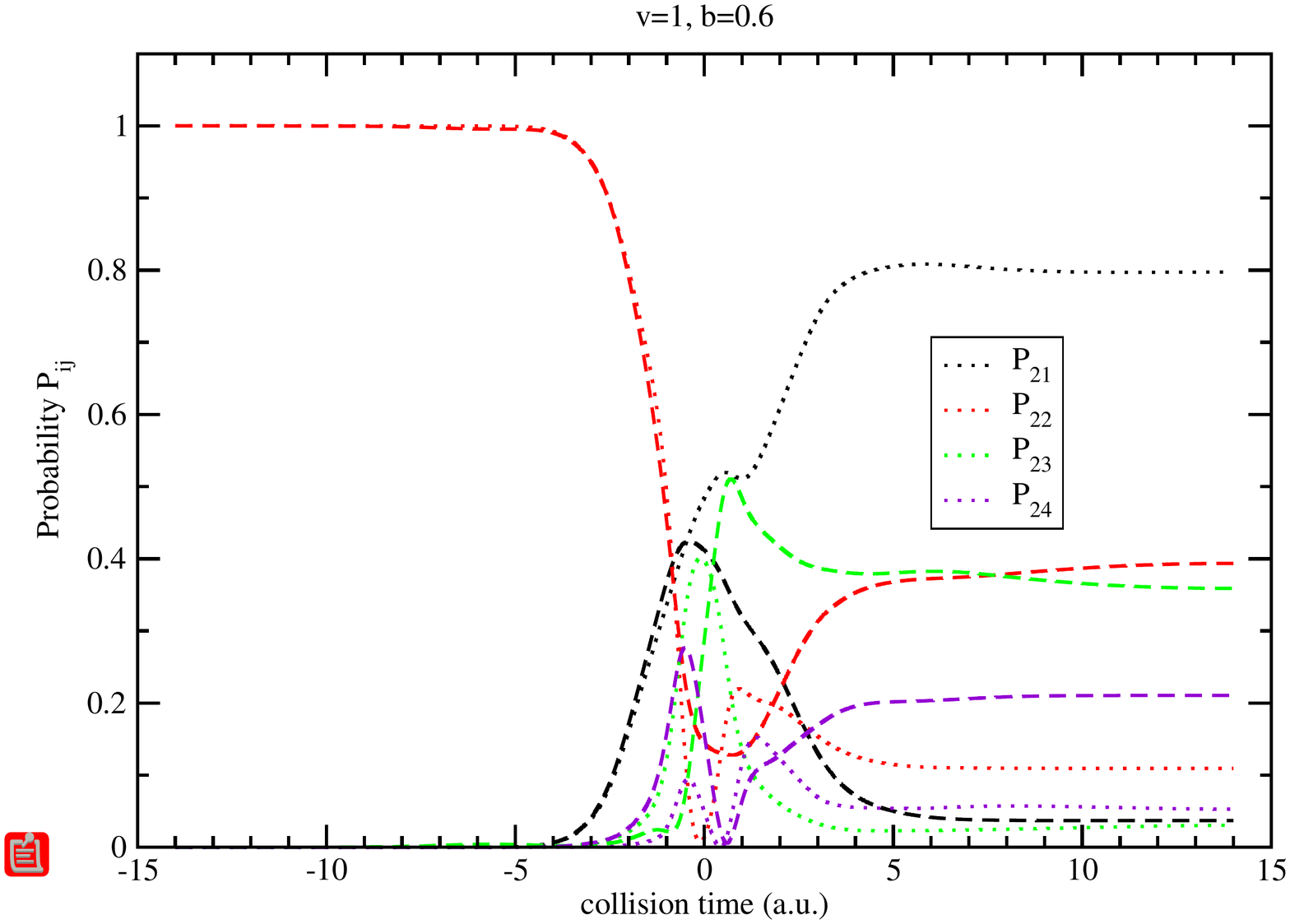}
\caption{Same as Fig.~\ref{probsihe3_4chn}, but comparing $n=4$ channels (dashed lines) and
$n=5$ channels (dotted lines) for $v_0=1.0$ a.u. and $b=0.6$ a$_0$.}
\label{probsihe3_5chnb}
\end{center}
\end{figure}




\begin{figure}[h]
\begin{center}
\includegraphics[scale=0.4]{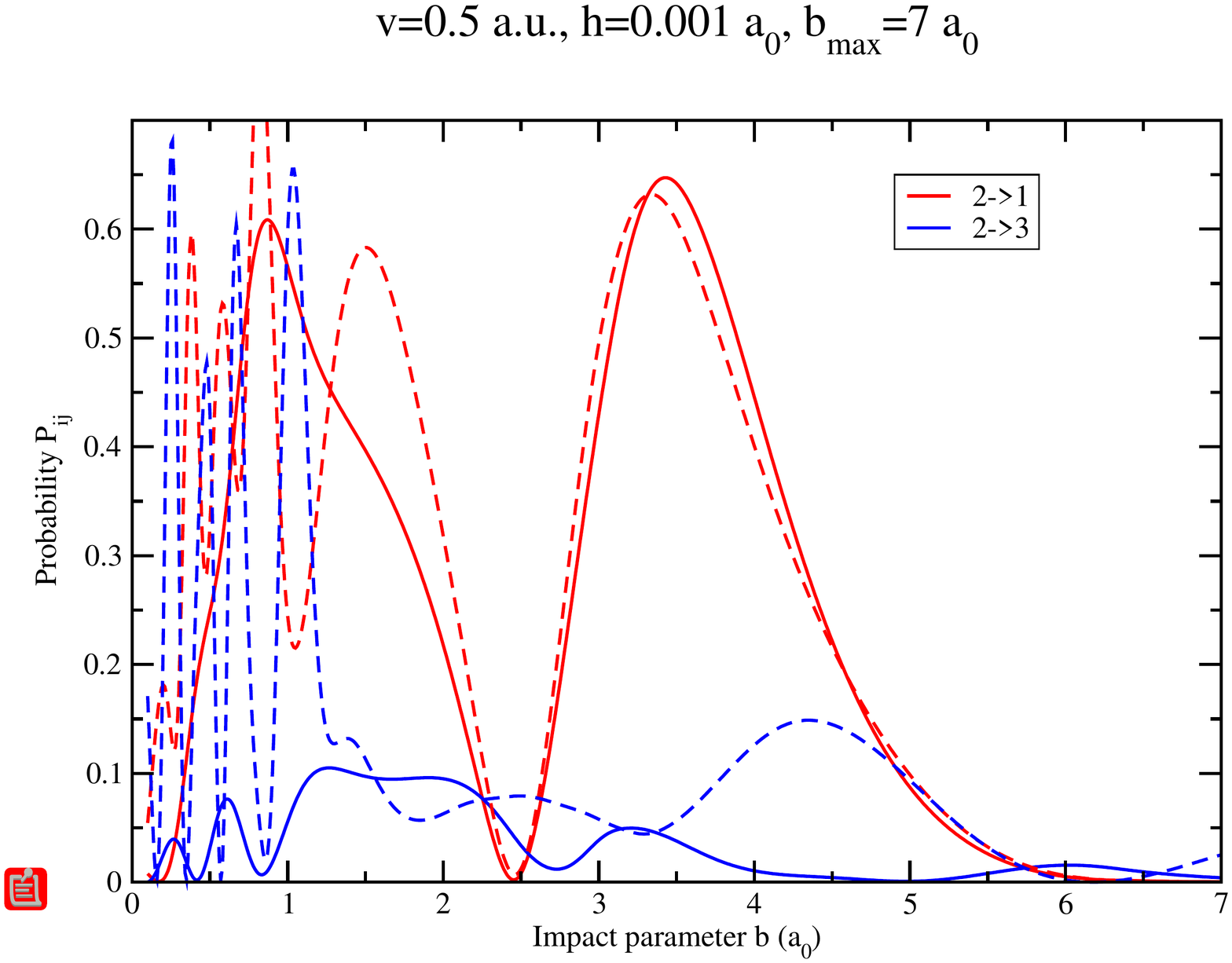}
\caption{Same as Fig. \ref{pbsihe3}, but comparing $n=3$ channels (solid lines) and
$n=5$ channel (dashed lines) for $v_0=0.5$ a.u.}
\label{pbsihe3_4chn}
\end{center}
\end{figure}



\begin{figure}[h]
\begin{center}
\includegraphics[scale=0.4]{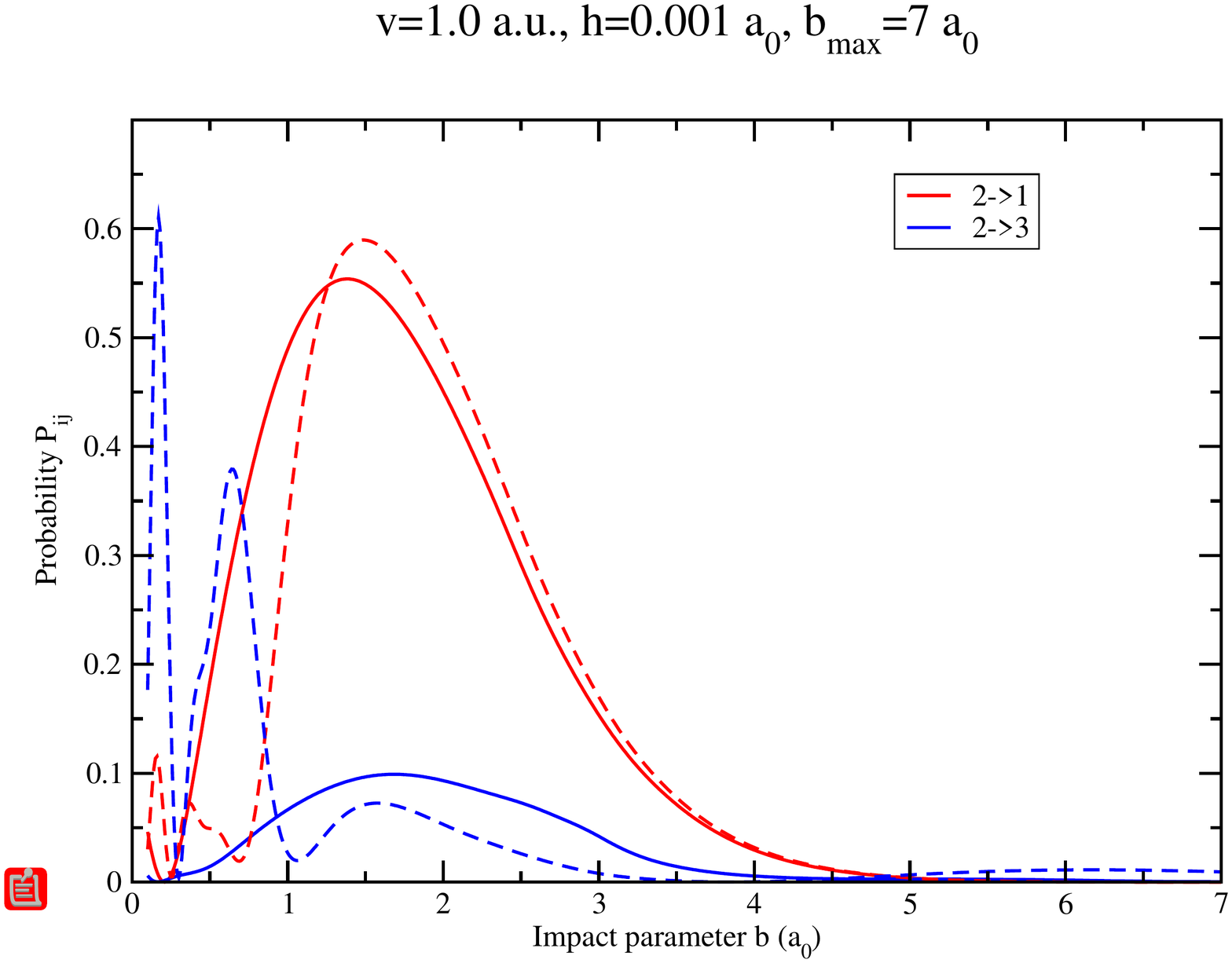}
\caption{Same as Fig.~\ref{pbsihe3_4chn}, but comparing $n=3$ channels (solid lines) and
$n=4$ channels (dashed lines) for $v_0=1.0$ a.u.}
\label{pbsihe3_4chnb}
\end{center}
\end{figure}

\begin{figure}[h]
\begin{center}
\includegraphics[scale=0.4]{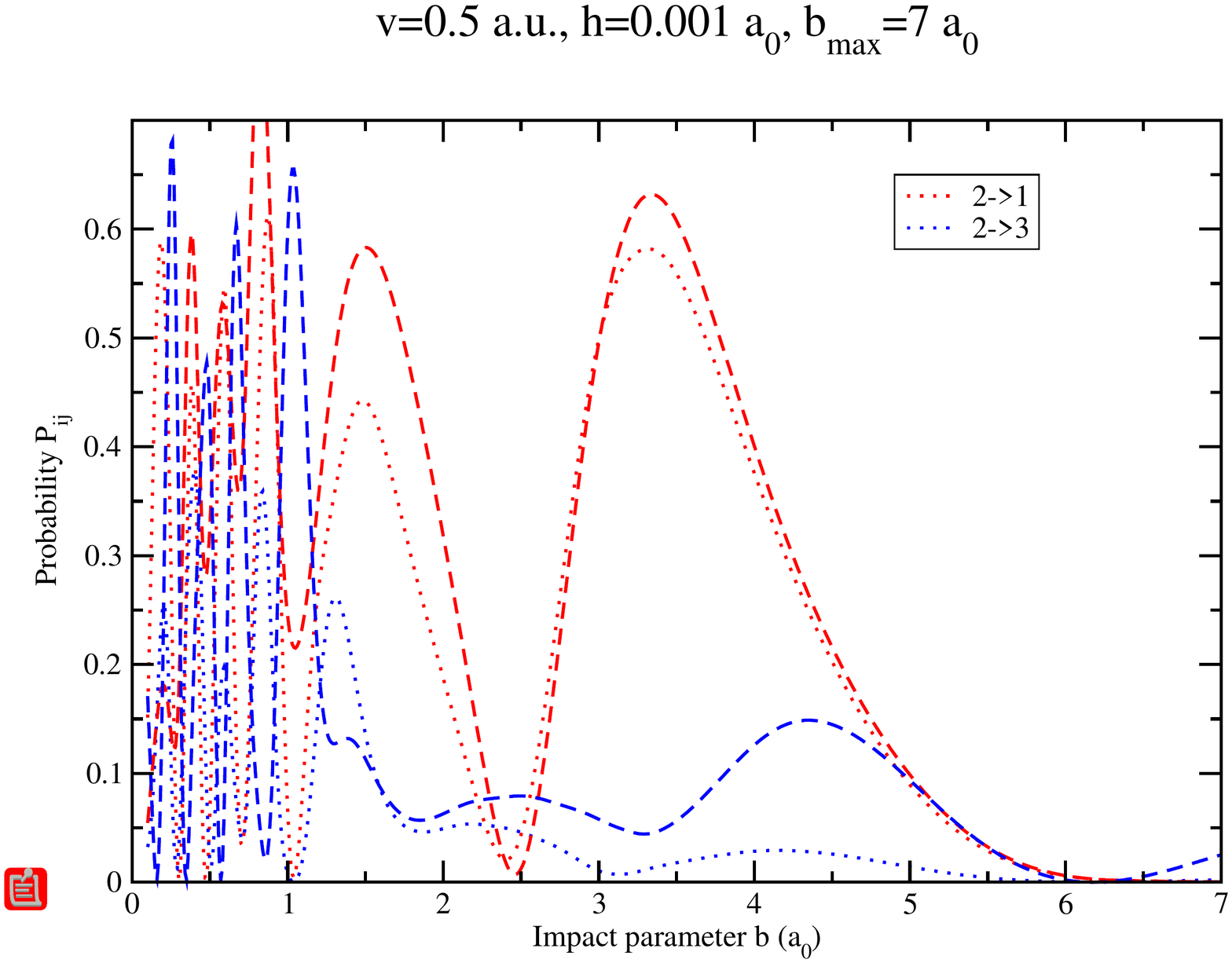}
\caption{Same as Fig.~\ref{pbsihe3_4chn}, but comparing $n=4$ channels (dashed lines) and
$n=5$ channels (dotted lines) for $v_0=0.5$ a.u.}
\label{pbsihe3_5chn}
\end{center}
\end{figure}

\begin{figure}[h]
\begin{center}
\includegraphics[scale=0.4]{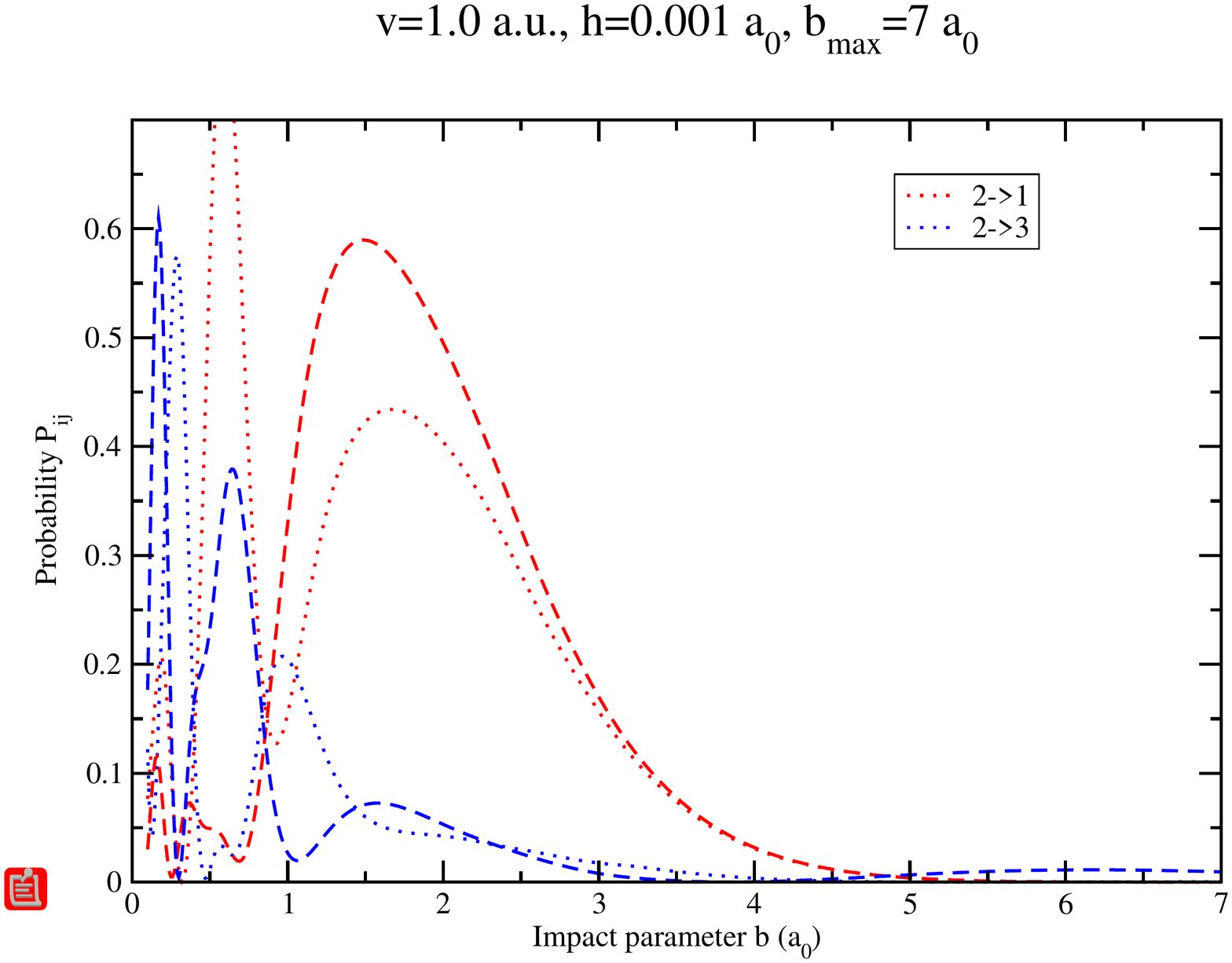}
\caption{Same as Fig.~\ref{pbsihe3_4chn}, but comparing $n=4$ channels (dashed lines) and
$n=5$ channels (dotted lines) for or $v_0=1.0$ a.u.}
\end{center}
\label{pbsihe3_5chnb}
\end{figure}

\begin{figure}[h]
\begin{center}
\includegraphics[scale=0.40]{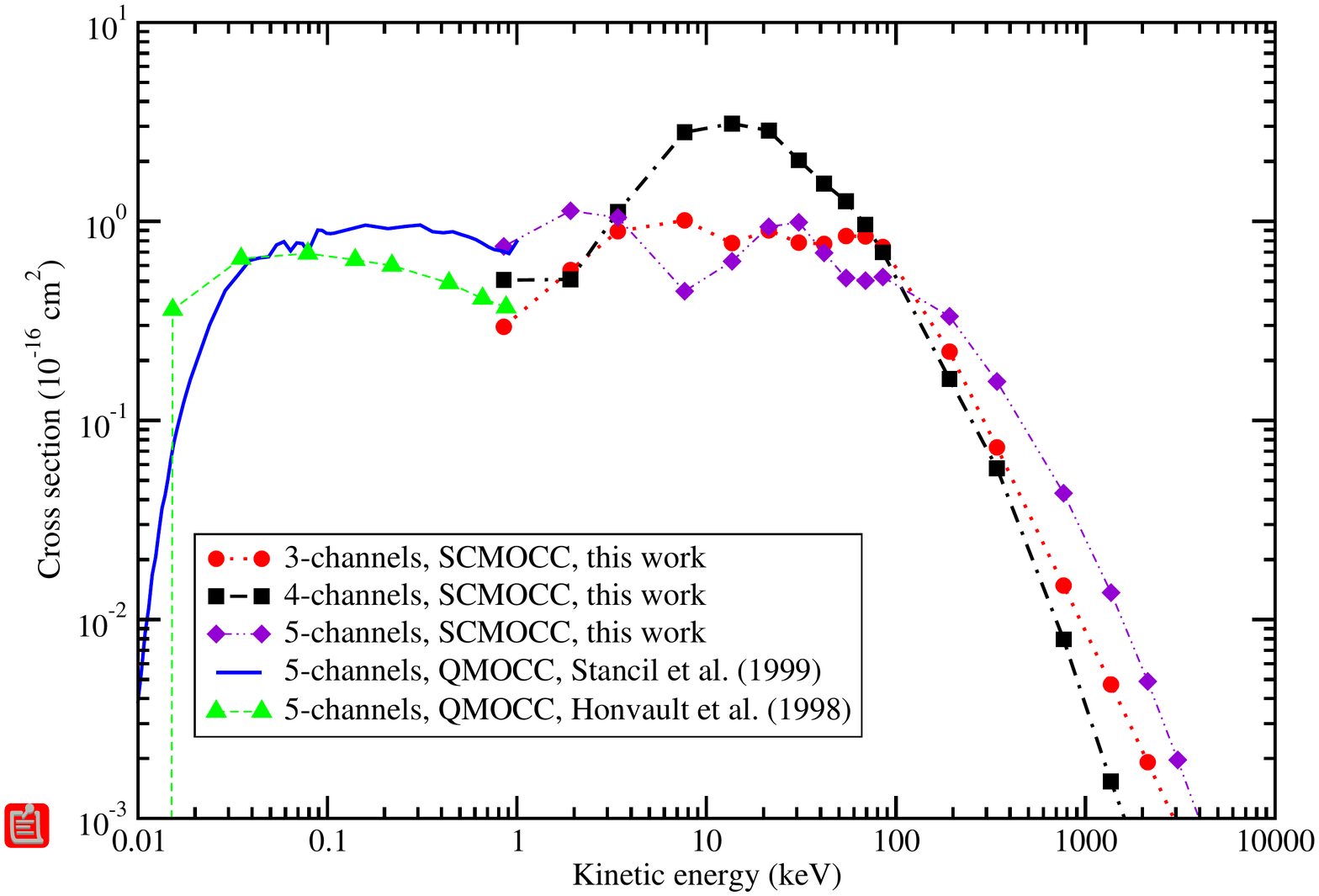}
\caption{The Si$^{3+}$ + He charge exchange cross section for the
$2\rightarrow 3$ transition comparing the current SCMOCC results to
earlier QMOCC results. Note the cross section is given as a function
of center-of-mass kinetic energy and the results of Ref.~\cite{sta99} used
the same diabatic potential as the current work.}
\label{cross3P5chn}
\end{center}
\end{figure}

\begin{figure}[h]
\begin{center}
\includegraphics[scale=0.40]{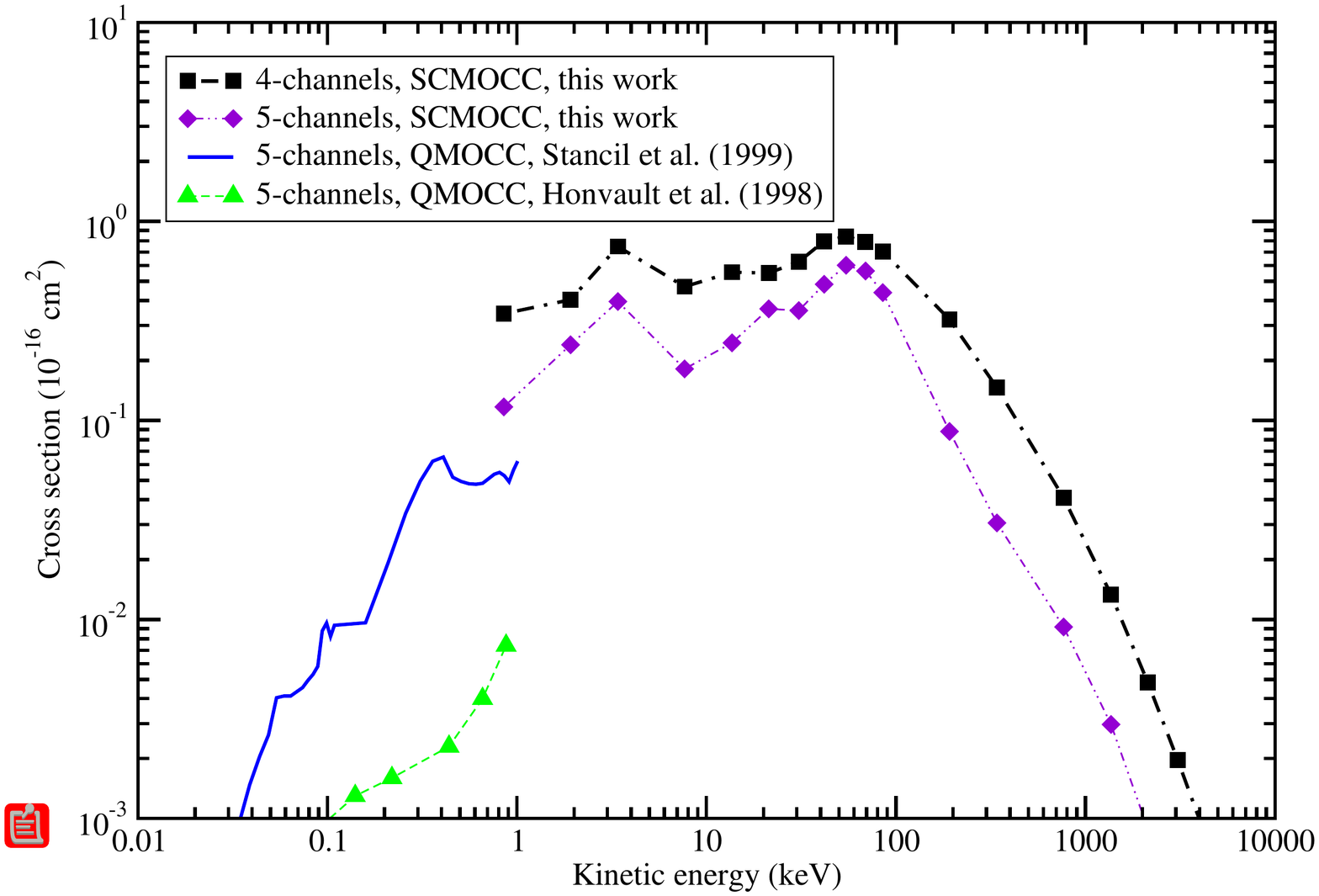}
\caption{The Si$^{3+}$ + He charge exchange cross section for the
$2\rightarrow 4$ transition comparing the current SCMOCC results to
earlier QMOCC results. Note the cross section is given as a function
of center-of-mass kinetic energy and the results of Ref.~\cite{sta99} used
the same diabatic potential as the current work.}
\label{cross1P}
\end{center}
\end{figure}

\begin{figure}[h]
\begin{center}
\includegraphics[scale=0.40]{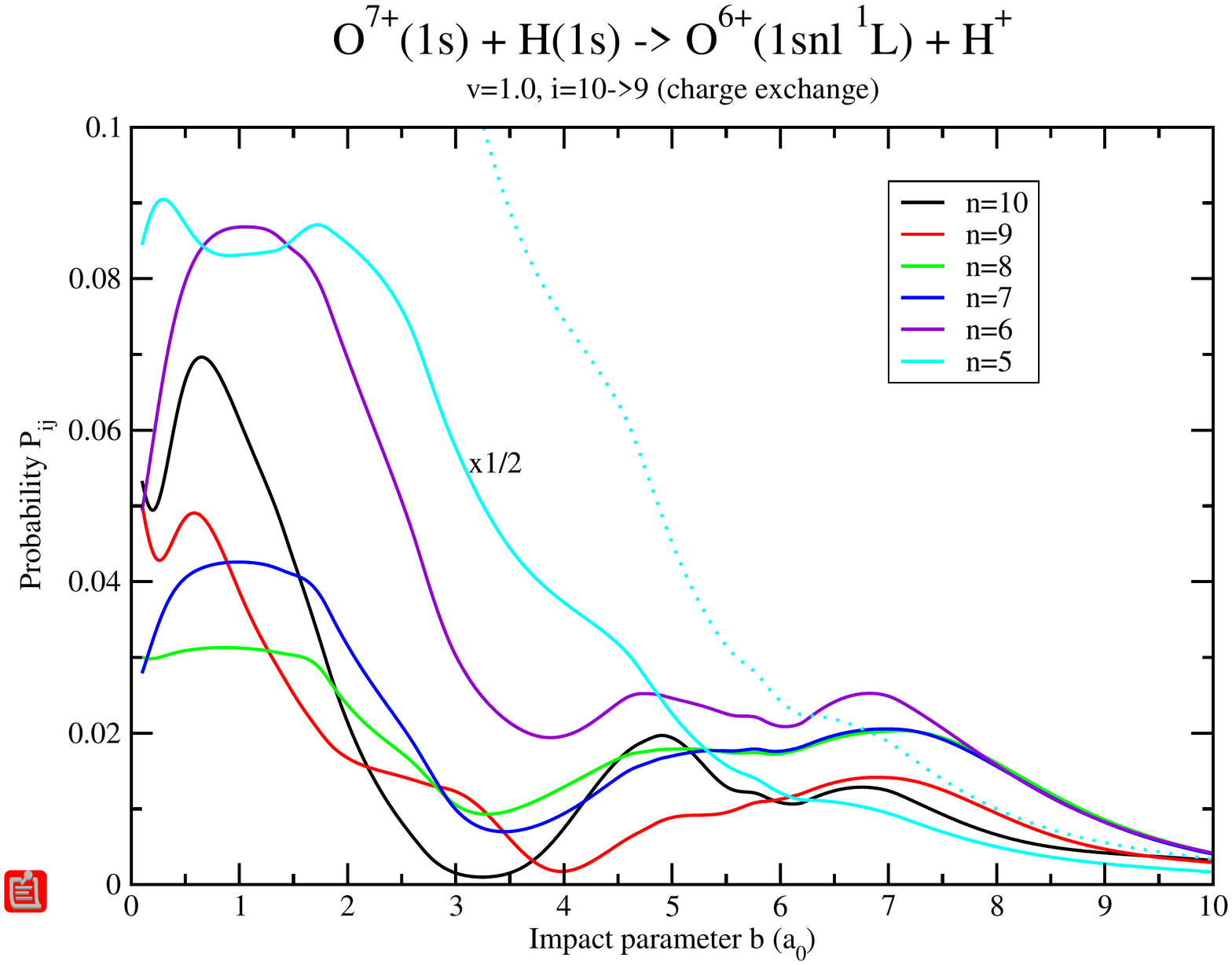}
\caption{The O$^{7+}$ + H probabilities versus impact parameter for the state-resolved charge
exchange reaction with product O$^{6+}$($1s5p~^1P$) + H$^+$ obtained from SCMOCC
calculations
with $n=5-10$ channels and $v=1.0$ a.u.}
\label{oh72}
\end{center}
\end{figure}

\section*{Acknowledgments}

This work was supported by the National Science Foundation under CDI grant
DMR-1029764. We thank Andrei  Galiautdinov, Joydip Ghosh, and Emily Pritchett
for helpful discussions.


\end{document}